\documentclass[reprint,superscriptaddress,amsmath,amsfonts,amssymb,aps,prb,showkeys]{revtex4-2}
\usepackage{tabularx}
\usepackage{graphicx}
\usepackage{dcolumn}
\usepackage{bm, color}
\usepackage{amsmath}
\usepackage{amsfonts}
\usepackage{amssymb}
\usepackage{amsmath}
\usepackage{here}
\usepackage{multirow}
\usepackage{ulem}
\usepackage{cancel}
\usepackage{bbm}
\providecommand{\U}[1]{\protect\rule{.1in}{.1in}}

\usepackage{hyperref}
\hypersetup{
	colorlinks = true,
}
\usepackage{physics}

\usepackage[whole]{bxcjkjatype}

\begin{document}

\title{Majorana-assisted nonlocal spin correlation in quasi-one-dimensional Kitaev spin liquids}

\author{Yuki Yamazaki}
\affiliation{RIKEN Cluster for Pioneering Research, Wako, Saitama 351-0198, Japan}
\affiliation{Department of Applied Physics, The University of Tokyo, Hongo, Tokyo, 113-8656, Japan}
\author{Shingo Kobayashi}
\affiliation{RIKEN Center for Emergent Matter Science, Wako, Saitama 351-0198, Japan}
\author{Akira Furusaki}
\affiliation{RIKEN Center for Emergent Matter Science, Wako, Saitama 351-0198, Japan}

\date{\today}

\begin{abstract} 
We propose Majorana-assisted nonlocal spin correlation as a manifestation of Majorana nonlocality in quasi-one-dimensional (1D) Kitaev spin liquids. Focusing on the flux-free sector of the Kitaev honeycomb model in a quasi-1D geometry, we uncover its topological nature and show that it hosts Majorana zero modes localized at both ends, which are stabilized by finite-size-induced topology. We further show that the nonlocal Majorana fermion parity operator, $P_{\text{MF}}=i\gamma_{\text{L}}\gamma_{\text{R}}$, is mapped to a nonlocal spin-string operator, producing an end-to-end spin correlation proportional to the product of $P_{\text{MF}}$ and total fermion parity operators when local perturbations remove redundant ground-state degeneracies while preserving the Majorana and total fermion parities in the flux-free sector. Numerical calculations confirm a finite nonlocal spin correlation generated by these Majorana zero modes without any local magnetization. Our results establish a concrete signature of intrinsic Majorana nonlocality in quantum spin liquids.
\end{abstract}

\maketitle

\begin{figure}[t]
    \centering
    \includegraphics[scale=0.565]{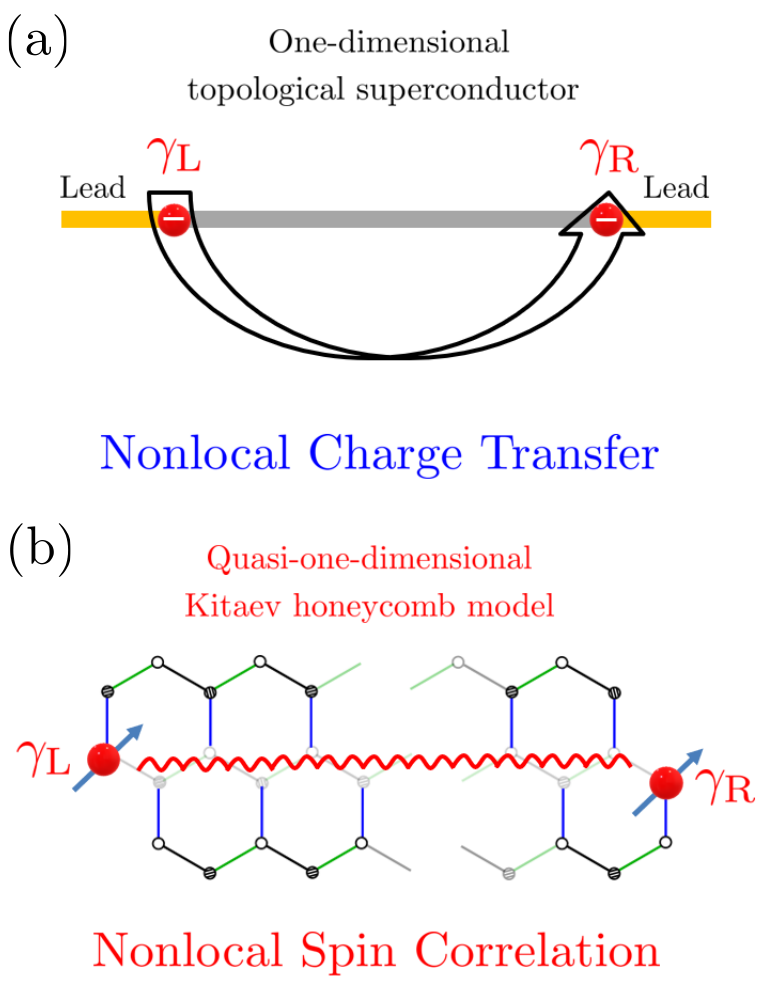}
    \caption{Sketch of Majorana-assisted nonlocal correlation. (a) nonlocal electric charge transfer in a one-dimensional (1D) topological superconductor and (b) nonlocal spin correlation in a quasi-1D Kitaev honeycomb model associated with Majorana zero modes, which are represented by $\gamma_{\text{L}}$ and $\gamma_{\text{R}}$.}
    \label{spin-teleportation}
\end{figure}

\section{Introduction}

Topological phases of matter have emerged as a central theme in modern condensed matter physics, offering platforms for realizing novel types of long-range entanglement and exotic excitations \cite{Hasan3045, Qi1057, Chiu16}. In particular, quantum spin liquids (QSLs) are exotic phases of matter characterized by topological order with fractionalized excitations \cite{Balents199, Savary016502, Zhou025003, Broholmaay0668, Knolle451, Wen165113, Wen041004}. Among the various models proposed to capture the essence of QSLs, the Kitaev honeycomb model  \cite{Kitaev2} holds a distinguished position due to its exact solvability and its realization of emergent Majorana fermions (MFs) as low-energy degrees of freedom.

The Kitaev honeycomb model supports a spin-liquid ground state that hosts both itinerant and localized MFs, which result in the fractional excitations of the spin-$1/2$ degrees of freedom. In particular, under an external magnetic field, the Kitaev honeycomb model enters a chiral spin-liquid phase that hosts topologically protected Majorana edge states and Majorana zero modes obeying the non-Abelian statistics \cite{Kitaev2}. Notably, recent thermal transport experiments on candidate Kitaev materials have reported a half-quantized thermal Hall effect \cite{Kasahara559, Yamashita220404, Yokoi568, Bruin401, Imamura3539}, 
reminiscent of the existence of Majorana chiral edge states in the chiral spin-liquid phase.

The nonlocal nature and non-Abelian statistics of Majorana zero modes have attracted great interest from both a fundamental and an application-oriented perspective, particularly in the context of fault-tolerant quantum computation \cite{Kitaev303, Nayak1083}. The braiding statistics of non-Abelian Majorana zero modes have been studied theoretically both in topological superconductors (TSCs) \cite{Alicea412, Posske023205} and in the Kitaev honeycomb model \cite{Jang085142, Harada241118, Harada214426}. Braiding protocols rely primarily on the topological properties of the excitations and are therefore largely insensitive to whether the underlying system carries charge degrees of freedom (as in TSCs) or not (as in the Kitaev honeycomb model).
In contrast, Majorana-assisted nonlocal charge transfer—a hallmark of Majorana nonlocality in mesoscopic TSCs—relies crucially on the presence of charge carriers. It enables an electron to tunnel nonlocally between spatially separated Majorana zero modes $\gamma_{\text{L}}, \gamma_{\text{R}}$, effectively transferring charge between two distant points with an amplitude that is independent of the system size \cite{Fu056402}; see Fig.~\ref{spin-teleportation} (a). This nonlocal tunneling process can serve as a sharp diagnostic of the nonlocal nature of Majorana zero modes.

This contrast naturally raises a compelling question:
Can the nonlocality of Majorana zero modes in the Kitaev honeycomb model manifest itself as a spin analogue of nonlocal charge transfer?
To give an answer to this question, in this work, we demonstrate that a nonlocal spin correlation can emerge between two spins on the opposite ends of a quasi-one-dimensional (quasi-1D) Kitaev honeycomb model, in a way that is independent of the system size (Fig.~\ref{spin-teleportation} (b)). The long-range spin correlation is shown to be mediated by Majorana zero modes and provides a manifestation of Majorana nonlocality unique to QSLs.

In the quasi-1D Kitaev honeycomb model, it is known that Majorana zero modes emerge under open boundary conditions (OBCs) \cite{Tadokoro011160}. 
We show that, in a quasi-1D topological phase, the boundary Majorana zero modes come in two distinct types depending on the anisotropy of the Kitaev interactions. One type is directly connected to the Majorana flat-band edge states of the two-dimensional (2D) Kitaev honeycomb model, whereas the other is generated in a gap caused by the overlap of the Majorana flat-band edge states. Both types of Majorana zero modes are stabilized by a 1D winding number of the quasi-1D model of itinerant Majorana fermions.
We reveal that the latter case is related to the finite-size-induced topology \cite{Fu096407, Linder205401, Liu041307, Lu115407, Sakamoto165432, Shan_2010, Zhang584, Rui1187485, EBIHARA2012885, Taskin066803, Ozawa045309, Yang246402, Wang064520, Asmar075419, Wu041014, Zhang137001, Chong6386, Cook045144, Flores125410, Adipta035146, ikegaya2025}. 

We then show that a finite nonlocal spin correlation between the edges in the quasi-1D Kitaev honeycomb model can arise in the flux-free ground-state sector, originating from the nonlocal Majorana fermion parity $i\gamma_{\text{L}}\gamma_{\text{R}}$ and its mapping to a nonlocal spin-string operator through the Jordan--Wigner transformation. Once redundant degeneracies from localized Majorana fermions (or $Z_2$ gauge fields) are lifted while a twofold Majorana-fermion-parity doublet remains, the system exhibits a size-independent nonlocal spin correlation without any net local magnetization. This result provides a concrete route to expose intrinsic Majorana nonlocality in Kitaev spin liquids and to probe nonlocal quantum correlations in QSLs \cite{Minakawa047204, Koga214421, Taguchi125139, Nasu024411, Misawa115117, Takahashi236701, Matsueda1yr5-sn98}.

The existence of Majorana zero modes at the ends of a quasi-1D Kitaev honeycomb model should be contrasted with the case of the $S=1$ Haldane chain \cite{HALDANE1983464,Haldane1153} and the AKLT model \cite{Affleck799,Affleck477}.
In the latter case the boundary $S=\frac12$ spins are free and uncorrelated, leading to four-fold degeneracy.
In our setting, the two-fold degenerate ground states of the quasi-1D Kitaev honeycomb model have two Majorana zero modes that are directly related to a string operator of spins connecting both ends, thereby causing a nonlocal spin correlation.

In previous studies of the Kitaev ladder \cite{Pedrocchi205412} and 2D honeycomb models with site vacancies under magnetic fields \cite{Takahashi236701}, the nonlocal spin correlation is characterized by Majorana zero modes localized at edges or around site vacancies.
However, our study differs from these works in two key aspects:
\begin{itemize}
\item[1.]  In the Kitaev ladder model \cite{Pedrocchi205412}, the nonlocal spin correlation in the ground states becomes finite only when the Majorana-fermion-parity degeneracy is lifted.
\item[2.] In the vacancy-based setups \cite{Takahashi236701}, each spin localized around site vacancies is polarized due to magnetic fields, and the observed nonlocal spin correlation contains this pre-determined alignment.
\end{itemize}
By contrast, in our study, there is no net magnetization and the ground states are {\it two}-fold degenerate due to the presence of the Majorana fermion parity $i\gamma_{\text{L}}\gamma_{\text{R}}$. Since each ground state has the same value of the nonlocal spin correlation between edge spins, the nonlocal spin correlation becomes finite even in the two-fold degenerate ground states.


This paper is organized as follows. Sec.~\ref{sec: Majorana_representation} introduces the Kitaev honeycomb model and our definition of the Jordan–Wigner transformation and Majorana representation (Sec.~\ref{sec: Jordan--Wigner}), then reviews the properties of the flux-free ground-state sector (Sec.~\ref{sec: ground} and Sec.~\ref{app:2dfluxfree}). Sec.~\ref{sec: topo_property} analyzes quasi-1D flux-free states and constructs a low-energy effective model for hybridized Majorana flat bands due to finite-size effects. Secs.~\ref{sec: Majorana zero modes} and~\ref{sec: Finite-size-effect-induced topological phase} clarify how system size and interaction anisotropy yield two distinct types of Majorana zero modes, one of which arises from finite-size-induced topology. Sec.~\ref{sec: nonlocal_spin_corr} formulates nonlocal spin correlations mediated by Majorana zero modes. Sec.~\ref{sec: numerics} presents the numerical demonstration of nonlocal spin correlation. We give technical details in the Appendices.

\section{Kitaev honeycomb model}\label{sec: Majorana_representation}

\subsection{Hamiltonian}\label{sec: Jordan--Wigner}

\begin{figure}[t]
    \centering
    \includegraphics[scale=0.315]{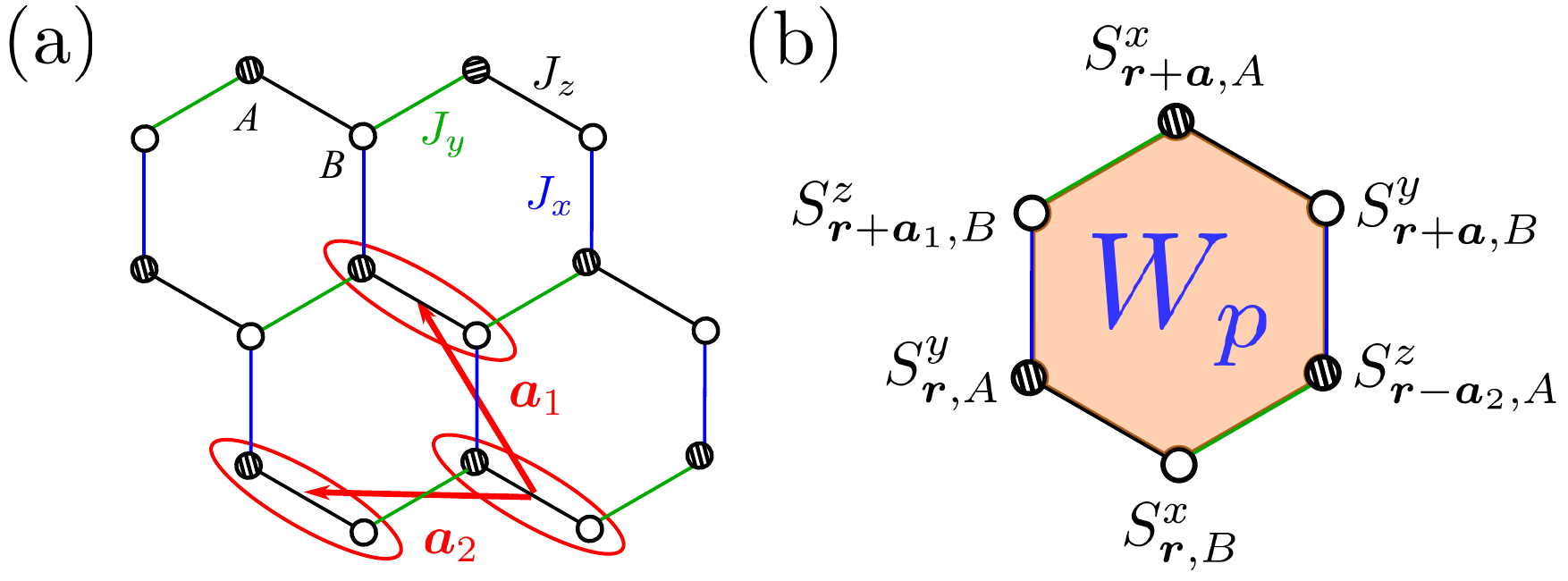}
    \caption{(a) Two-dimensional honeycomb lattice structure of the Kitaev honeycomb model. (b) Schematic illustration of $\mathbb{Z}_2$ flux of hexagonal plaquette $W_p$.}
    \label{Kitaev-picture}
\end{figure}

The Kitaev honeycomb model~\cite{Kitaev2} is a quantum spin model defined on a
two-dimensional honeycomb lattice with localized spin-$1/2$ magnetic moments and bond-dependent anisotropic
interactions. The exchange interactions are all of Ising type, where the interacting spin component depends on the three types of nearest-neighbor bonds on the tricoordinate structure. The Hamiltonian is given by
\begin{align}
H &= -\sum_{\mu=x,y,z}J_{\mu}\sum_{\langle i,j \rangle_{\mu}}S^{\mu}_i S^{\mu}_j, \\ \nonumber
  &= -\sum_{\bm{r}}\Big(J_{x}S^{x}_{\bm{r}-\bm{a}_1,A} S^{x}_{\bm{r},B} +J_{y}S^{y}_{\bm{r}-\bm{a}_2,A} S^{y}_{\bm{r},B}\\ &\hspace{39mm} +J_{z}S^{z}_{\bm{r},A} S^{z}_{\bm{r},B}\Big), 
\label{Kitaev}
\end{align}
where $J_{\mu}$ is the exchange coupling constant on the $\mu$ bonds ($J_{\mu}>0$), and $S_i^{\mu}$ is the $\mu$ component of the spin-$1/2$ operator at site $i$. The summation of $\langle i,j \rangle_{\mu}$ is taken over nearest-neighbor spin pairs. In the second line, we introduce the lattice vector as $\bm{r} = r_{\bm{a}_1}\bm{a}_1 + r_{\bm{a}_2}\bm{a}_2$, where $\bm{a}_1=(\frac{1}{2},\frac{\sqrt{3}}{2}) $ and $ \bm{a}_2=(-\frac{1}{2},\frac{\sqrt{3}}{2})$ denote unit vectors of the honeycomb lattice, $A$ and $B$ label two sublattices in a unit cell, and $r_{\bm{a}_1},r_{\bm{a}_2} \in \mathbb{Z}$ describe the position of the unit cell. The schematic picture is shown in Fig.~\ref{Kitaev-picture} (a). 
   
The ground state of the Kitaev honeycomb model is exactly obtained by introducing a Majorana representation of the spin operators~\cite{Kitaev2}. Here, we employ the Majorana representation through the Jordan--Wigner transformation introduced in Refs.~\cite{Chen193101, Feng087204, Chen075001, Motome012002}. In the Jordan--Wigner transformation, the spin-$1/2$ operators are rewritten in terms of spinless fermion operators as
\begin{align}
\nonumber
S^{+}_j &= (S^{-}_j)^{\dagger} = S^{x}_j+iS^{y}_j = a^{\dagger}_j\prod^{j-1}_{j'=1}(1-2n_{j'}), \\
S^{z}_j &= n_j - \frac{1}{2}, \label{eq:JW}
\end{align}
where $a^{\dagger}_j$ and $a_j$ are the creation and annihilation operators for the spinless fermions, respectively, and $n_j=a^{\dagger}_j a_j$ is the number operator. $a^{\dagger}_j$ and $a_j$ satisfy the anticommutation relations as $\{a_j^{\dagger}, a_k\}= \delta_{jk}$, $\{a_j^{\dagger},a_k^{\dagger}\}=0$, and  $\{a_j,a_k\}=0$, where $\delta_{jk}$ is the Kronecker delta. The subscript $j=1,2,\cdots$ labels the lattice sites, which are ordered according to the sequence used in performing the Jordan--Wigner transformation; namely, the system is regarded as a one-dimensional chain composed of the $x$ and $y$ bonds. Examples of the labeling are shown in the Appendix~\ref{app: Jordan--Wigner}.
Substituting Eq.~(\ref{eq:JW}) into Eq.~(\ref{Kitaev}), the Hamiltonian is transformed into 
\begin{align}
\nonumber
H = &\ \frac{J_{x}}{4}\sum_{\bm{r}} \left(a_{\bm{r},B} - a^{\dagger}_{\bm{r},B} \right)\left(a_{\bm{r}-\bm{a}_1,A}+a^{\dagger}_{\bm{r}-\bm{a}_1,A} \right) \\ \nonumber  &+ \frac{J_{y}}{4}\sum_{\bm{r}} \left(a_{\bm{r},B} - a^{\dagger}_{\bm{r},B} \right)\left(a_{\bm{r}-\bm{a}_2,A}+a^{\dagger}_{\bm{r}-\bm{a}_2,A} \right) \\ 
&-\frac{J_{z}}{4}\sum_{\bm{r}} \left(2n_{\bm{r},B}-1\right)\left(2n_{\bm{r},A}-1 \right),
\end{align}
where each term is described by the product of fermions at the A and B sites due to the bipartite structure.
The Majorana representation is obtained by writing the spinless fermion operators with Majorana fermion operators. The relation is given by 
\begin{align}
\nonumber
\gamma_{\bm{r},A} = a_{\bm{r},A}+a^{\dagger}_{\bm{r},A}, \quad \bar{\gamma}_{\bm{r},A} = -i(a_{\bm{r},A}-a^{\dagger}_{\bm{r},A}), \\
\gamma_{\bm{r},B} = -i(a_{\bm{r},B}-a^{\dagger}_{\bm{r},B}), \quad \bar{\gamma}_{\bm{r},B} = a_{\bm{r},B}+a^{\dagger}_{\bm{r},B}, 
\label{fermion-to-Majorana}
\end{align}
where $\gamma_{\bm{r},s}$ and $\bar{\gamma}_{\bm{r},s}$ are the so-called itinerant and localized Majorana fermion operators. The Majorana fermion operators satisfy $\gamma^{\dagger}_{\bm{r},s}=\gamma_{\bm{r},s}$ and $\{\gamma_{\bm{r},s},\gamma_{\bm{r}',s'}\}=2\delta_{\bm{r}\bm{r}'}\delta_{ss'}$. $\bar{\gamma}_{\bm{r},s}$ also satisfy the same relations. $\gamma_{\bm{r},s}$ and $\bar{\gamma}_{\bm{r},s}$ anticommute with each other.
Using Eq.~(\ref{fermion-to-Majorana}), the Hamiltonian is finally rewritten into
\begin{align}
\nonumber
H = \frac{i}{4}\sum_{\bm{r}}\big(J_x \gamma_{\bm{r},B}\gamma_{\bm{r}-\bm{a}_1,A} + J_y \gamma_{\bm{r},B}\gamma_{\bm{r}-\bm{a}_2,A} \\ \hspace{29.8mm} + J_z \eta_{\bm{r}} \gamma_{\bm{r},B}\gamma_{\bm{r},A}\big),
\label{Majorana-Kitaev}
\end{align}
where $\eta_{\bm{r}}\equiv i\bar{\gamma}_{\bm{r},A} \bar{\gamma}_{\bm{r},B}$ is defined on the $z$ bond and satisfies
\begin{align}
[H,\eta_{\bm{r}}]=0, \quad \eta_{\bm{r}}^2 =1, \quad [\eta_{\bm{r}},\eta_{\bm{r}'}] =0,
\end{align}
Therefore, each $\eta_{\bm{r}}$ takes $\pm 1$ and is a conserved quantity. Note that $\eta_{\bm{r}}$ is related to another conserved quantity, referred to as the $\mathbb{Z}_2$ flux $W_p$. The $\mathbb{Z}_2$ flux is defined for each hexagonal plaquette on the honeycomb lattice as
\begin{align}
    W_p = \sigma^x_{\bm{r},B}\sigma^y_{\bm{r},A}\sigma^z_{\bm{r}+\bm{a}_1,B}\sigma^x_{\bm{r}+\bm{a},A}\sigma^y_{\bm{r}+\bm{a},B}\sigma^z_{\bm{r}-\bm{a}_2,A},
\end{align}
where the product is taken for the six sites on the plaquette $p$ in a clockwise manner. Here $\sigma_j^{\mu} = 2 S_j^{\mu}$ is the $\mu$th component of the Pauli matrices, and $\bm{a}=\bm{a}_1-\bm{a}_2$. The hexagonal plaquette $W_p$ is schematically shown in Fig.~\ref{Kitaev-picture} (b). Using Eqs.~(\ref{eq:JW}) and (\ref{fermion-to-Majorana}), $W_p$ is also expressed as
\begin{align}
W_p = \prod_{\bm{r}\in p} \eta_{\bm{r}}, \label{eq:defwp}
\end{align}
where the product runs over the two $z$ bonds belonging to plaquette $p$.

To facilitate the subsequent discussion, we summarize the relation between the spin operators and the Majorana fermion operators as follows.
\begin{align}
\nonumber
&\sigma^z_{\bm{r},A} = i\gamma_{\bm{r},A}\bar{\gamma}_{\bm{r},A}, \quad \sigma^z_{\bm{r},B} = i\bar{\gamma}_{\bm{r},B}\gamma_{\bm{r},B}, \\ \nonumber
&\sigma^y_{\bm{r},A} = -\bar{\gamma}_{\bm{r},A}\prod^{j<(\bm{r},A)}_{j=1}(-\sigma^z_j), \quad \sigma^y_{\bm{r},B} = -\gamma_{\bm{r},B}\prod^{j<(\bm{r},B)}_{j=1}(-\sigma^z_j), \\
&\sigma^x_{\bm{r},A} = \gamma_{\bm{r},A}\prod^{j<(\bm{r},A)}_{j=1}(-\sigma^z_j), \quad \sigma^x_{\bm{r},B} =\bar{\gamma}_{\bm{r},B}\prod^{j<(\bm{r},B)}_{j=1}(-\sigma^z_j),
\label{Spin-to-Majorana}
\end{align}
where $j<(\bm{r},s), \ s=A,B$ denotes sites whose Jordan–Wigner label is smaller than that of $(\bm{r},s)$.

\subsection{Quantum spin-liquid ground state}\label{sec: ground}
We examine the connection between the ground state of Eq.~(\ref{Majorana-Kitaev}) and the conserved variables $\eta_{\bm{r}}$ by analyzing the dimension of the Hamiltonian. Let $N$ denote the number of sites. The dimension of the Hilbert space of the original spin Hamiltonian~(\ref{Kitaev}) is $2^{N}$. In deriving the Majorana representation, each spin degree of freedom is replaced by two Majorana fermion degrees of freedom. Thus, $2^{N}$ is decomposed into the number of itinerant and localized Majorana fermions as
\begin{align}
2^N = \underbrace{2^\frac{N}{2}}_{\gamma} \times \underbrace{2^\frac{N}{2}}_{\bar{\gamma}}. \label{dimension}
\end{align}
Eq.~(\ref{Majorana-Kitaev}) indicates that the original spin model can be mapped onto a system of non-interacting itinerant Majorana fermions coupled to $\eta_{\bm{r}}$ or $W_p$. Hence, the Hamiltonian can be expressed in a block-diagonal form, with each block specified by a configuration of $\{\eta_{\bm{r}}\}$ or $\{W_p\}$; the total Hamiltonian with the dimension $2^N$ is decomposed into a direct sum of the sectors specified by $\{W_p\}$. The number of $\{W_p\}$ configurations depends on the boundary condition as follows.

First, we consider systems with the periodic boundary condition (PBC) imposed in the $\bm{a}_1$ and $\bm{a}_2$ directions, referred to as two-dimensional (2D) systems in this paper. Under the PBC, the number of $\{W_p\}$ configurations is $2^{N/2}$. Thus, the decomposition can be rewritten as 
\begin{align}
2^N = \underbrace{2^\frac{N}{2}}_{\gamma} \times \underbrace{2^\frac{N}{2}}_{W_p}.
\label{dimension-periodic}
\end{align}
Comparing Eq.~(\ref{dimension}) with Eq.~(\ref{dimension-periodic}), the degrees of freedom of $\bar{\gamma}$ is equivalent to  the number of $\{W_p\}$ configurations. Once the configuration of $\{W_p\}$ is specified, the remaining degrees of freedom correspond to itinerant Majorana fermions $\gamma$, and the problem reduces to that of non-interacting fermions. The problem is further simplified by using Lieb's theorem~\cite{Lieb2158}, which tells that the flux configuration with all $W_p=+1$ gives the lowest energy state. This configuration is referred to as the flux-free state. Note that Lieb's theorem is applicable only when at least two of three $J_\mu$ are equal. It does not apply to the cases with generic $J_{\mu}$. Nevertheless, by comparing the energies for other configurations of $\{W_p\}$, the flux-free state is deduced to be the ground state in the entire parameter space of $J_{\mu}$~\cite{Kitaev2}. 

The flux-free state is a QSL with short-range spin correlations~\cite{Baskaran247201}
\begin{align}
&\langle S^\mu_{i}S^\nu_{j}\rangle  \neq 0 \ \  \text{for} \ \ \mu=\nu \ \text{ and } \ i,j \in \langle i,j\rangle_{\mu}, \label{eq:spin-corrv1}\\
&\langle S^\mu_{i}S^\nu_{j}\rangle =  0 \ \  \text{for the others}. \label{eq:spin-corrv2}
\end{align}
Namely, the spin correlations have a nonzero value only for the $\mu$ component on the nearest-neighbor $\mu$ bonds under PBCs.

\begin{figure}[t]
    \centering
    \includegraphics[scale=0.31]{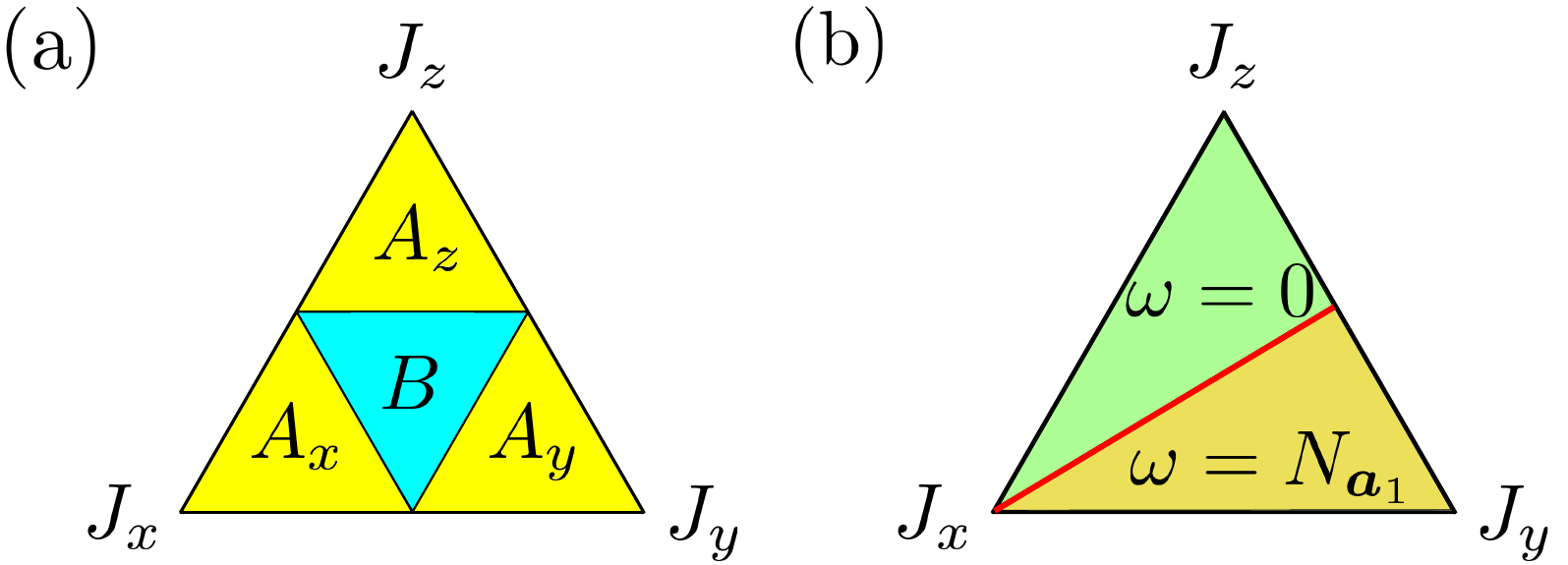}
    \caption{Phase diagram with $J_x+J_y+J_z=\text{constant}$ in (a) two-dimensional Kitaev honeycomb model~\cite{Kitaev2} and (b) quasi-1D Kitaev honeycomb model with the open boundary condition in the $\bm{a}_1$ direction~\cite{Tadokoro011160}. } 
    \label{Fig: Kitaev-phase-diagram}
\end{figure}

\subsection{Two-dimensional flux-free states}
\label{app:2dfluxfree}
Here, we briefly review the flux-free states of the two-dimensional Kitaev honeycomb model, with an emphasis on the relation to the 1D topological invariants. We impose the PBC in both the $\bm{a}_1$ and $\bm{a}_2$ directions and choose all $\eta_{\bm{r}}$ to be $+1$.
Let $N_{\bm{a}_{1}}$ and $N_{\bm{a}_{2}}$ denote the lattice size along the $\bm{a}_{1}$ and $\bm{a}_{2}$ directions.
By substituting the Fourier transformation
\begin{equation}
\gamma_{\bm{r},s} = \gamma^{\dagger}_{\bm{r},s} =  \sqrt{\frac{2}{N_{\bm{a}_1}N_{\bm{a}_2}}}\sum_{\bm{k}}e^{i\bm{k}\cdot \bm{r}}\gamma_{\bm{k},s}
\quad(s=A,B)
\end{equation}
to Eq.~(\ref{Majorana-Kitaev}), the Hamiltonian is rewritten as
\begin{align}
H  = \frac{1}{2}\sum_{\bm{k}} \bm{\gamma}^{\dagger}_{\bm{k}} 
H(\bm{k})
\bm{\gamma}_{\bm{k}},
\label{Kitaev-Hamiltonian}
\end{align}
with 
\begin{align}
H(\bm{k}) &= \frac{1}{2}\left(
\begin{array}{cc}
  0   & q(\bm{k})  \\
  q^\ast(\bm{k})  &  0
\end{array}
\right), \label{2D-Kitaev-Hamiltonian} \\
\bm{\gamma}_{\bm{k}}&=(\gamma_{\bm{k},A},\gamma_{\bm{k},B})^{\text{T}},
\end{align}
where $q(\bm{k}) \equiv -iJ_x e^{i\bm{k}_{\bm{a}_1}}-iJ_y e^{i\bm{k}_{\bm{a}_2}} - iJ_z$. The Majorana operators satisfy $\gamma^{\dagger}_{\bm{k},s} = \gamma_{-\bm{k},s} \ (s=A,B)$ and $\{\gamma^{\dagger}_{\bm{k},s},\gamma_{\bm{k}',s'}\}=\delta_{\bm{k}\bm{k}'}\delta_{ss'} \ (s,s'=A,B)$. 
The energy spectrum of $H(\bm{k})$ is given by 
\begin{align}
\left|E(\bm{k})\right| = \frac{1}{2}\left|J_x e^{i\bm{k}_{\bm{a}_1}} + J_y e^{i\bm{k}_{\bm{a}_2}} + J_z\right|.
\end{align}
When $J_x=J_y=J_z$, $\left|E(\bm{k})\right|=0$ at $K=(\frac{4\pi}{3},\frac{2\pi}{3})$ and $K'=(\frac{2\pi}{3},\frac{4\pi}{3})$ points.
As shown in Fig.~\ref{Fig: Kitaev-phase-diagram} (a), the flux-free states are categorized into four phases: the $B$ phase represents a gapless phase ($J_{\alpha}\le J_{\beta}+J_{\gamma}, \ \{\alpha,\beta,\gamma\}=\{x,y,z\}$), while the $A_{\alpha}$ $(\alpha=x,y,z)$ phases are gapped phases ($J_{\alpha} > J_{\beta}+J_{\gamma}$ for the $A_{\alpha}$ phase).

\begin{table}[t]
	\caption{1D topological invariants~(\ref{winding-number}) in each phase. The entry ``1" indicates that there exist some $k_{\bm{a}_j}$ at which $W_{\bm{a}_i}(k_{\bm{a}_j})$ takes a nonzero value, while ``0" indicates that $W_{\bm{a}_i}$ vanishes for any $k_{\bm{a}_j}$ ($i\ne j$).}
	\begin{tabular}{ccccc}
		\hline\hline
		 & $B$ & $A_x$ & $A_y$ & $A_z$
		\\
		\hline
  	$|W_{\bm{a}_1}|$ & $1$ & $1$ & $0$ & $0$
		\\
		$|W_{\bm{a}_2}|$ & $1$ & $0$ & $1$ & $0$
		\\
		\hline\hline
	\end{tabular}
	\label{Tab: winding-number}
\end{table}

The Hamiltonian in Eq.~(\ref{Kitaev-Hamiltonian}) preserves the time-reversal ($\Theta$), particle-hole ($C$), and chiral ($\Gamma$) symmetries as 
\begin{align}
\Theta H(\bm{k}) \Theta^{-1} &= H(-\bm{k}), \ \Theta = \tau_z K, \\
C H(\bm{k}) C^{-1} &= -H(-\bm{k}), \ C =K, \\
\Gamma H(\bm{k}) \Gamma^{-1} &= -H(\bm{k}), \ \Gamma=\tau_z,
\end{align}
where $\tau_i$'s ($i=x,y,z$) denote the Pauli matrices for the sublattice degrees of freedom. The above equations mean that $H(\bm{k})$ belongs to class BDI in the Altland--Zirnbauer symmetry classification~\cite{Altland1142}. Thus, we can define a 1D topological invariant,
\begin{align}
W_{\bm{a}_i} (k_{\bm{a}_j}) = \frac{1}{2\pi}\int^{\pi}_{-\pi} dk_{\bm{a}_i} \frac{\partial}{\partial k_{\bm{a}_i}} \text{arg}[q(k_{\bm{a}_i},k_{\bm{a}_j})] \label{winding-number}
\end{align}
where $i,j=1,2 \ (i\neq j)$.
The 1D topological invariants calculated in each phase are shown in Table~\ref{Tab: winding-number}.

In the $B$ phase, $W_{\bm{a}_1}$ and $W_{\bm{a}_2}$ are nonzero at $k_{\bm{a}_j}$ between the $K$ and $K'$ points.
As is well known, the $B$ phase realizes a Majorana Chern insulator with a nonzero Chern number under a magnetic field~\cite{Kitaev2}.

In the $A_x$ ($A_y$) phase, $W_{\bm{a}_1}$  ($W_{\bm{a}_2}$) is nonzero at arbitrary $k_{\bm{a}_2}$ ($k_{\bm{a}_1}$), which implies that a Majorana flat band is formed at the edges along the $\bm{a}_2$ ($\bm{a}_1$) direction~\cite{Thakurathi235434, Mizoguchi184418}.
In the $A_z$ phase both $W_{\bm{a}_1}$ and $W_{\bm{a}_2}$ vanish because the $J_z$ interaction couples itinerant Majorana fermions within each unit cell.

\section{Quasi-one-dimensional flux-free states}\label{sec: topo_property}

\subsection{Quantum spin-liquid ground state in quasi-1D system}\label{sec: ground-quasi}

Next, we consider quasi-1D systems with $N_{\bm{a}_1}<N_{\bm{a}_2}$, imposing the open boundary condition (OBC) in the $\bm{a}_1$ direction while keeping the PBC in the $\bm{a}_2$ direction. In this setting, the localized Majorana fermions $\bar{\gamma}$ located at sites on the edges are not included in the $\mathbb{Z}_2$ flux operator $W_p$ of any plaquette. Thus, the total number of $\bar{\gamma}_i$ generally does not coincide with the number of allowed $\mathbb{Z}_2$ flux configurations $\{W_p\}$. Let $D$ denote the number of redundant binary degrees of freedom due to this mismatch. The decomposition becomes
\begin{align}
2^N = \underbrace{2^\frac{N}{2}}_{\gamma} \times \underbrace{2^{N_{W_p}}}_{W_p} \times \underbrace{2^{D}}_{\bar{\gamma}},
\label{dimension-open}
\end{align}
where the last factor $2^D$ comes from the localized Majorana fermions $\bar{\gamma}$ that are not included in any $W_p$.
The number of hexagonal plaquettes is denoted by $N_{W_p}$, which depends on the lattice geometry of quasi-1D systems.
Thus, $D$ is given by
\begin{align}
D= \frac{N}{2} - N_{W_p}.
\label{Eq:extra-degeneracy}
\end{align}
Therefore, even after fixing the $W_p$ configuration, a $2^D$-fold degeneracy remains. 

In quasi-1D systems, Lieb's theorem remains applicable when reflection symmetry is preserved~\cite{Lieb2158}. In general situations, numerical evidence suggests that the flux-free sector has lower energy than other flux configurations~\cite{Mizoguchi184418}. Therefore, unless otherwise stated, we focus on the flux-free states in the following discussion. The short-range spin correlation relations~(\ref{eq:spin-corrv1}) and (\ref{eq:spin-corrv2}) hold in the bulk of the flux-free states in quasi-1D systems. However, situations can be different at the edges. We will show in the Sec.\ \ref{sec: nonlocal_spin_corr} that a nonlocal spin correlation between edges emerge when Majorana zero modes are present.

In the remainder of Sec.~\ref{sec: topo_property}  and in Sec.~\ref{sec: nonlocal_spin_corr}, we focus on the topological properties of itinerant Majorana fermions $\gamma$ under the flux-free condition, ignoring the redundant $2^D$-fold degeneracy.

\subsection{Topological phases}
We discuss topologically-protected Majorana zero modes in the quasi-1D Kitaev honeycomb model under the OBC in the $\bm{a}_1$ direction and the PBC in the $\bm{a}_2$ direction. By applying the Fourier transformation 
\begin{equation}
\bm{\gamma}_{\bm{r},s} = \sqrt{\frac{2}{N_{\bm{a}_2}}}\sum_{k_{\bm{a}_2}=1} e^{ik_{\bm{a}_2}r_{\bm{a}_2}} \bm{\gamma}_{r_{\bm{a}_1},s}(k_{\bm{a}_2})
\quad(s=A,B)
\end{equation}
to Eq.~(\ref{Majorana-Kitaev}), the Hamiltonian is rewritten as
\begin{align}
\nonumber
&H=\frac{1}{2}\sum_{k_{\bm{a}_2}}\bm{\gamma}^{\dagger}_{k_{\bm{a}_2}}H(k_{\bm{a}_2})\bm{\gamma}_{k_{\bm{a}_2}}, \\ \nonumber
&\bm{\gamma}_{k_{\bm{a}_2}} = [\gamma_{1,A}(k_{\bm{a}_2}),\gamma_{1,B}(k_{\bm{a}_2}), \\&\hspace{20mm}\cdots,\gamma_{N_{\bm{a}_1},A}(k_{\bm{a}_2}),\gamma_{N_{\bm{a}_1},B}(k_{\bm{a}_2})]^{\text{T}}, \label{tight-binding}
\end{align}
where $H(k_{\bm{a}_2})$ is $2N_{\bm{a}_1} \times 2N_{\bm{a}_1} $ Hermitian matrix. The elements of the matrix are given by
\begin{align}
&[H(k_{\bm{a}_2})]_{(j-1,A),(j,B)} = [H(k_{\bm{a}_2})]^{*}_{(j,B),(j-1,A)} =\frac{J_x}{2i},\\
&[H(k_{\bm{a}_2})]_{(j,A),(j,B)} = [H(k_{\bm{a}_2})]^{*}_{(j,B),(j,A)}  =\frac{J_ye^{ik_{\bm{a}_2}}+J_z}{2i},
\end{align}
where $j = 1,2, \cdots, N_{\bm{a}_1}$ and the other elements are zero.
The Fourier-transformed Majorana fermion operators satisfy 
$\gamma^{\dagger}_{r_{\bm{a}_1},s}(k_{\bm{a}_2}) = \gamma_{r_{\bm{a}_1},s}(-k_{\bm{a}_2})$ and the anticommutation relation $\{\gamma^{\dagger}_{r_{\bm{a}_1},s}(k_{\bm{a}_2}),\gamma_{r'_{\bm{a}_1},s'}(k'_{\bm{a}_2})\}=\delta_{k_{\bm{a}_2}k'_{\bm{a}_2}}\delta_{r_{\bm{a}_1}r'_{\bm{a}_1}}\delta_{ss'}$ $(s,s'=A,B)$.

The Hamiltonian in~Eq.~(\ref{tight-binding}) satisfies the time-reversal ($\Theta$), particle-hole (C), and chiral ($\Gamma$) symmetries:
\begin{align}
\Theta H(k_{\bm{a}_2}) \Theta^{-1} &= H(-k_{\bm{a}_2}), \ \Theta = \mathbbm{1}_{N_{\bm{a}_1}} \otimes \tau_z K, \\
C H(k_{\bm{a}_2}) C^{-1} &= -H(-k_{\bm{a}_2}), \ C = \mathbbm{1}_{2N_{\bm{a}_1}} K, \\
\Gamma H(k_{\bm{a}_2}) \Gamma^{-1} &= -H(k_{\bm{a}_2}), \ \Gamma=\mathbbm{1}_{N_{\bm{a}_1}} \otimes \tau_z,
\end{align}
where $\mathbbm{1}_{N}$ is a $N \times N$ identity matrix. Since $C^2=\mathbbm{1}_{2N_{\bm{a}_1}}$ and $\Theta^2=\mathbbm{1}_{2N_{\bm{a}_1}}$, the Hamiltonian belongs to class BDI. In this class, there are 1D topological phases~\cite{schnyder, kitaevtopo, Schnyder09, ryu} classified by the $\mathbb{Z}$-valued topological invariant
\begin{align}
\omega = \frac{1}{2\pi i}\int^{\pi}_{-\pi} dk_{\bm{a}_2} \frac{\partial}{\partial k_{\bm{a}_2}} \text{ln det}[h(k_{\bm{a}_2})], \label{winding-number2}
\end{align}
where the matrix $h(k_{\bm{a}_2})$ is defined by
\begin{align}
U^{\dagger}H(k_{\bm{a}_2})U &\equiv \left[
\begin{array}{cc}
0 & h(k_{\bm{a}_2})\\
h^{\dagger}(k_{\bm{a}_2}) & 0
\end{array}
\right], \\ \nonumber
U^{\dagger}\bm{\gamma}_{k_{\bm{a}_2}} &= [\gamma_{1,A}(k_{\bm{a}_2}),\cdots, \gamma_{N_{\bm{a}_1},A}(k_{\bm{a}_2}), \\
&\quad \quad  \gamma_{1,B}(k_{\bm{a}_2}),\cdots, \gamma_{N_{\bm{a}_1},B}(k_{\bm{a}_2})]^{\text{T}}, \nonumber
\end{align}
with a unitary matrix $U$.
By numerically calculating Eq.~(\ref{winding-number2}) as a function of $J_x$, $J_y$, and $J_z$, the topological phase diagram of flux-free states in the quasi-1D Kitaev honeycomb model is obtained as shown in Fig.~\ref{Fig: Kitaev-phase-diagram} (b)~\cite{Tadokoro011160}. There are two phases: a trivial phase ($\omega=0$) for $J_y<J_z$ and a topological phase ($|\omega| = N_{\bm{a}_1}$) for $J_y>J_z$. According to the bulk-edge correspondence, $N_{\bm{a}_1}$ Majorana zero modes per edge are expected to emerge in the topological phase when the OBC is additionally imposed in the $\bm{a}_2$ direction. 

In the following, we show that reducing the system size of the 2D Kitaev honeycomb model to a quasi-1D geometry can generate an effective 1D topology inherited from the hybridized Majorana flat-band states. This finite-size-induced topology protects Majorana zero modes at edges in a quasi-1D system.

\begin{figure}[thb]
    \centering
    \includegraphics[scale=0.15]{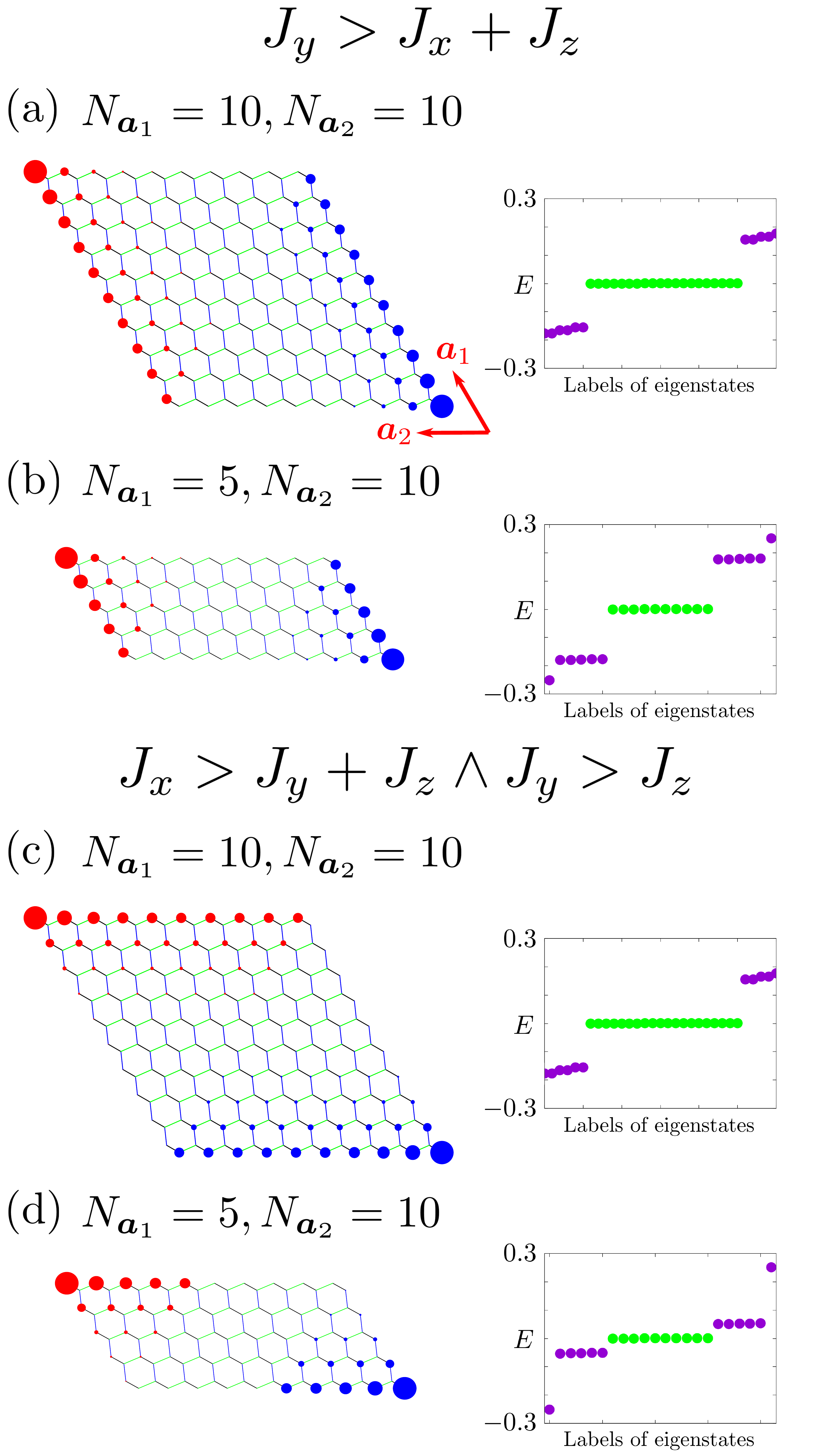}
    \caption{Figures show the probability density of Majorana zero modes (left) and the corresponding energy spectrum (right) for the systems with the full open boundary condition, where the parameters are set to $J_{y}=1.7$, $J_{x}=1.29$, $J_{z}=0.01$ for (a) and (b) and $J_{x}=1.7$, $J_{y}=1.29$, $J_{z}=0.01$ for (c) and (d). The lattice sizes are $N_{\bm{a}_2} =10$, $N_{\bm{a}_1} =10$ for (a) and (c) and $N_{\bm{a}_2} =10$, $N_{\bm{a}_1} =5$ for (b) and (d). In the probability-density plots, the red and blue filled circles denote the $A$ and $B$ sublattices, respectively, and the circle size is proportional to $\rho_{\ell}$ defined in Eq.~(\ref{eq:edge_density}). In the energy spectra, the purple and green dots correspond to excited states in the bulk and boundary zero modes, respectively. There are $N_{\rm MF}=20$ green dots in (a) and (c), and 10 green dots in (b) and (d).}
    \label{Fig: quasi-0D-wavefunction}
\end{figure}

\subsection{Majorana modes}\label{sec: Majorana zero modes}

To visualize the spatial profile of the Majorana zero modes, we numerically diagonalize Eq.~(\ref{Majorana-Kitaev}) with the OBCs for both $\bm{a}_1$ and $\bm{a}_2$ directions. In what follows, we fix $N_{\bm{a}_2}=10$ and compare two widths, $N_{\bm{a}_1}=10$ (2D-like) and $N_{\bm{a}_1}=5$ (quasi-1D), as shown in Fig.~\ref{Fig: quasi-0D-wavefunction}. Let 
\begin{align}
\bm{u}^{(\alpha)}\equiv\left(u^{(\alpha)}_{1,1,A},u^{(\alpha)}_{1,1,B},...,u^{(\alpha)}_{N_{\bm{a}_1},N_{\bm{a}_2},A},u^{(\alpha)}_{N_{\bm{a}_1},N_{\bm{a}_2},B}\right),
\end{align}
be the normalized eigenvectors associated with Majorana zero modes ($\alpha=1,2,...,N_{\text{MF}}$), where $N_{\text{MF}}$ is the total number of Majorana zero modes that are highlighted by green dots in the right panels of Fig.~\ref{Fig: quasi-0D-wavefunction}.
The quantity plotted by red and blue circles on the left panels of Fig.~\ref{Fig: quasi-0D-wavefunction} is the 
probability density of these Majorana zero modes,
\begin{align}
\rho_{\ell} \equiv \sum^{N_{\text{MF}}}_{\alpha=1} \left|u^{(\alpha)}_{\ell}\right|^2,
\label{eq:edge_density}
\end{align}
where $\ell \equiv(\bm{r},s), \ s=A,B$ and an overall normalization factor has been scaled for visualization purposes in Fig.~\ref{Fig: quasi-0D-wavefunction}. Through the spatial profile of the Majorana zero modes in Eq.~(\ref{eq:edge_density}), we verify their existence and localization at the edges and discuss their dependence on the relative strength of $J_x$ and $J_y$ by comparing two regimes: (i) $J_y > J_x+J_z$ and (ii) $J_x > J_y + J_z\, \land \, J_y>J_z$.

\subsubsection{\texorpdfstring{$J_y > J_x+J_z$}{J_y > J_x+J_z}}

Fig.~\ref{Fig: quasi-0D-wavefunction} (a) and (b) show the numerical results for $J_y > J_x+J_z$.
This parameter regime represents both the $A_y$ phase in the 2D phase diagram [Fig.~\ref{Fig: Kitaev-phase-diagram}(a)] and the quasi-1D topological phase [Fig.~\ref{Fig: Kitaev-phase-diagram}(b)].

For $N_{\bm{a}_1} = N_{\bm{a}_2} = 10$ [Fig.~\ref{Fig: quasi-0D-wavefunction} (a)], the system shows a characteristic behavior of the $A_y$ phase in the 2D Kitaev honeycomb model, having a Majorana flat band (i.e., zero modes) at the edges along the $\bm{a}_1$ direction.
As we discussed in the previous section, the $A_y$ phase is characterized by the winding number $W_{\bm{a}_2}$ defined in Eq.~(\ref{winding-number}), which equals 1 for every $k_{\bm{a}_1}$.
Since $W_{\bm{a}_2}$ is defined at each $k_{\bm{a}_1}$ point, the number of Majorana zero modes at an edge along the $\bm{a}_1$ direction is given by $N_{\bm{a}_1}$.

In the quasi-1D geometry [Fig.~\ref{Fig: quasi-0D-wavefunction} (b)] where $N_{\bm{a}_1}$ is reduced to $N_{\bm{a}_1}=5$, we observe $N_{\bm{a}_1}$ Majorana zero modes localized at each edge along the $\bm{a}_1$ direction.
This is consistent with the nontrivial value of the invariant, $|\omega|=N_{\bm{a}_1}$ in the quasi-1D topological phase.
Since the number of Majorana zero modes is $N_{\bm{a}_1}$ both in the 2D and quasi-1D geometries, the zero modes in Fig.~\ref{Fig: quasi-0D-wavefunction} (b) can also be regarded as a Majorana flat band in the 2D $A_y$ phase, protected by $W_{\bm{a}_2}$ with $N_{\bm{a}_1}=5$.

\subsubsection{\texorpdfstring{$J_x > J_y+J_z \land J_y>J_z$}{J_x > J_y+J_z \land J_y>J_z}}

Figures~\ref{Fig: quasi-0D-wavefunction} (c) and (d) show the numerical results for the parameter regime where $J_x > J_y+J_z$ and $J_y>J_z$. This regime corresponds to the 2D $A_x$ phase and the quasi-1D topological phase in Figs.~\ref{Fig: Kitaev-phase-diagram} (a) and (b), respectively.

For $N_{\bm{a}_1} = N_{\bm{a}_2} = 10$ [Fig.~\ref{Fig: quasi-0D-wavefunction} (c)], the system exhibits the characteristic behavior of the $A_x$ phase in the 2D Kitaev honeycomb model, hosting a Majorana flat band (zero modes) protected by the nontrivial winding number $W_{\bm{a}_1}=1$ at the edges along the $\bm{a}_2$ direction.
Since $W_{\bm{a}_1}$ is defined at each $k_{\bm{a}_2}$ point, the number of Majorana zero modes per edge is $N_{\bm{a}_2}$.

In the quasi-1D geometry [Fig.~\ref{Fig: quasi-0D-wavefunction} (d)] where $N_{\bm{a}_1}$ is reduced to $N_{\bm{a}_1} = 5$, each edge along the $\bm{a}_2$ direction has $N_{\bm{a}_1}$ Majorana zero modes that are dictated by the quasi-1D winding number $\omega$, which is equal to $|\omega|=N_{\bm{a}_1}$. Note that this is different from the number of zero modes in the 2D $A_x$ phase, $N_{\bm{a}_2}$.
Fig.~\ref{Fig: quasi-0D-wavefunction} (d) shows that the Majorana zero modes reside on the edges along the $\bm{a}_2$ direction, as in the 2D geometry [Fig.~\ref{Fig: quasi-0D-wavefunction} (c)], with the probability density decaying rapidly from a corner.
This suggests that the Majorana zero modes in Fig.~\ref{Fig: quasi-0D-wavefunction} (d) should be originated from the Majorana flat-band states of the 2D geometry upon reducing the system width $N_{\bm{a}_1}$.
In the next section, we show that this behavior is captured by a finite-size-induced topology \cite{Yang246402, Asmar075419, Zhang137001, Chong6386, Cook045144, Flores125410, Adipta035146, ikegaya2025} associated with the low-energy subspace of the hybridized Majorana flat-band states.

\subsubsection{Comparison of energy spectra in two regimes}

The difference between the two parameter regimes can also be seen in the energy spectra shown in the right panels of Fig.~\ref{Fig: quasi-0D-wavefunction}. For $J_y > J_x+J_z$, isolated Majorana zero modes (i.e., a Majorana flat band) appear within the bulk energy gap as shown in Fig.~\ref{Fig: quasi-0D-wavefunction} (a) and (b), where the number of Majorana zero modes increases with increasing $N_{\bm{a}_1}$.
Under the condition $J_x > J_y + J_z \land J_y>J_z$, a Majorana flat band is also formed at $N_{\bm{a}_1}=N_{\bm{a}_2}$ in Fig.\ \ref{Fig: quasi-0D-wavefunction} (c), which shows exactly the same energy spectrum as in Fig.~\ref{Fig: quasi-0D-wavefunction} (a).
In fact, the system studied in Fig.~\ref{Fig: quasi-0D-wavefunction}(a) is related to (c) by mirror reflection that exchanges the directions $\bm{a}_1$ and $\bm{a}_2$.
Finally, as $N_{\bm{a}_1}$ decreases, the Majorana flat-band states of opposite edges hybridize and open a mini-gap in the Majorana flat band. In Fig.~\ref{Fig: quasi-0D-wavefunction} (d), we find that $2N_{\bm{a}_1}$ Majorana zero modes remain within this mini-gap.

\begin{table}[t]
\centering
\caption{Summary of the winding numbers used in this paper. Each invariant is defined for a class-BDI Hamiltonian
with chiral symmetry, and protects Majorana zero modes localized at edges parallel to the $\bm{a}_1$ or $\bm{a}_2$ direction.}
\begin{tabular}{ccl}
\hline\hline
Invariant & Eq. & \ \ Geometry and Hamiltonian  \\
\hline
$W_{\bm{a}_i}(k_{\bm{a}_j})$ &
(\ref{winding-number}) &
\ \ 2D Hamiltonian $H(\bm{k})$ in Eq.~(\ref{2D-Kitaev-Hamiltonian}) \\
$\omega$ &
(\ref{winding-number2}) &
\ \ Q1D Hamiltonian $H(k_{\bm{a}_2})$ in Eq.~(\ref{tight-binding}) \\
$\widetilde{W}_{\bm{a}_2}$ &
(\ref{winding-number3}) &
\ \ Effective Hamiltonian $\widetilde{H}(k_{\bm{a}_2})$ in Eq.~(\ref{Sizeeffect-effective-Hamiltonian})
\\
\hline\hline
\end{tabular}
\label{Tab:winding-summary}
\end{table}

\subsection{Finite-size-induced topological phase}\label{sec: Finite-size-effect-induced topological phase}

Figure~\ref{Fig: quasi-0D-wavefunction} (d) indicates that Majorana flat-band states residing opposite edges hybridize and open a finite-size gap, leaving Majorana zero modes in this gap. In the following, we discuss the topological origin of these Majorana zero modes using an effective Hamiltonian that describes the hybridization of the Majorana flat-band states localized at opposite edges. The low-energy effective Hamiltonian is given by
\begin{align}
\widetilde{H}(k_{\bm{a}_2}) &= \left[
\begin{array}{cc}
0 & \tilde{q}(k_{\bm{a}_2}) \\
\tilde{q}^{*}(k_{\bm{a}_2})  & 0
\end{array}
\right],
\label{Sizeeffect-effective-Hamiltonian}
\end{align}
with
\begin{align}
    &\tilde{q}(k_{\bm{a}_2})=\frac{iJ_x}{2}[X (k_{\bm{a}_2})]^2 [\zeta(k_{\bm{a}_2})]^{N_{\bm{a}_1}}, \label{eq:tildeq}\\  
&X(k_{\bm{a}_2})  = \sqrt{\frac{1-|\zeta(k_{\bm{a}_2})|^2}{1-|\zeta(k_{\bm{a}_2})|^{2N_{\bm{a}_1}}}}, \label{eq:Xfn}\\
&\zeta(k_{\bm{a}_2}) = -\frac{J_y e^{ik_{\bm{a}_2}}+J_z}{J_x}, \label{eq:zetafn}
\end{align}
the derivation of which is given in the Appendix~\ref{app: finite-size}. The off-diagonal component $\tilde{q}(k_{\bm{a}_2})$ describes the hybridization between the Majorana flat bands.
The effective Hamiltonian $\widetilde{H}$ has the energy eigenvalues $\pm |\tilde{q}|$, which vanish at 
$k_{\bm{a}_2}=\pm \pi$ when $J_y=J_z$, i.e., at the topological phase boundary in Fig.~\ref{Fig: Kitaev-phase-diagram} (b).

We examine the topology of the effective Hamiltonian~(\ref{Sizeeffect-effective-Hamiltonian}), which shares the same symmetries with the original Hamiltonian and therefore belongs to class BDI. Thus, we can define a 1D topological invariant
\begin{align}
\widetilde{W}_{\bm{a}_2} &= \frac{1}{2\pi}\int^{\pi}_{-\pi} dk_{\bm{a}_2} \frac{\partial}{\partial k_{\bm{a}_2}} \text{arg}[\tilde{q}(k_{\bm{a}_2})], \label{winding-number3}
\end{align}
which takes the following values:
\begin{align}
|\widetilde{W}_{\bm{a}_2}| = \begin{cases}N_{\bm{a}_1}  &\text{when $\ |J_y|>|J_z|$,} \\  0  &\text{when $\ |J_z|>|J_y|$.} \end{cases}
\label{Sizeeffect-winding-number}
\end{align}
The winding number $\widetilde{W}_{\bm{a}_2}$ is equivalent to $\omega$.
However, $\widetilde{W}_{\bm{a}_2}$ is defined from the effective Hamiltonian $\widetilde{H}$, while $\omega$ is defined from the original quasi-1D Hamiltonian $H(k_{\bm{a}_2})$.
We summarize the definition of winding numbers and their defining Hamiltonian in Table~\ref{Tab:winding-summary}.

The analysis based on the effective Hamiltonian shows that the Majorana zero modes in Fig.~\ref{Fig: quasi-0D-wavefunction}(d) are boundary states of the Majorana flat-band states that carry a nontrivial topological invariant $\widetilde{W}_{\bm{a}_2}$. In other words, these Majorana zero modes exist as a consequence of the nontrivial topology induced by finite-size effects~\cite{Yang246402, Asmar075419, Zhang137001, Chong6386, Cook045144, Flores125410, Adipta035146, ikegaya2025}.

In the quasi-1D geometry with $J_y>J_z$, the gap remains open as the relative strength between $J_x$ and $J_y$ is changed from the regime $J_x > J_y$ [Fig.~\ref{Fig: quasi-0D-wavefunction} (d)] to the regime $J_y > J_x$ [Fig.~\ref{Fig: quasi-0D-wavefunction} (b)], indicating a continuous crossover within the quasi-1D topological phase in Fig.~\ref{Fig: Kitaev-phase-diagram}.

\subsection{Edge termination}
Finally, we discuss the effect of edge terminations. When the edges are constructed so as to preserve the unit cells, the number of Majorana zero modes coincides with the 1D topological invariant~(\ref{winding-number3}). In contrast, when choosing an edge termination that breaks a unit cell, the number of Majorana zero modes deviates from the 1D topological invariant by the number of broken unit cells.

\begin{figure}[t]
    \centering
    \includegraphics[scale=0.12]{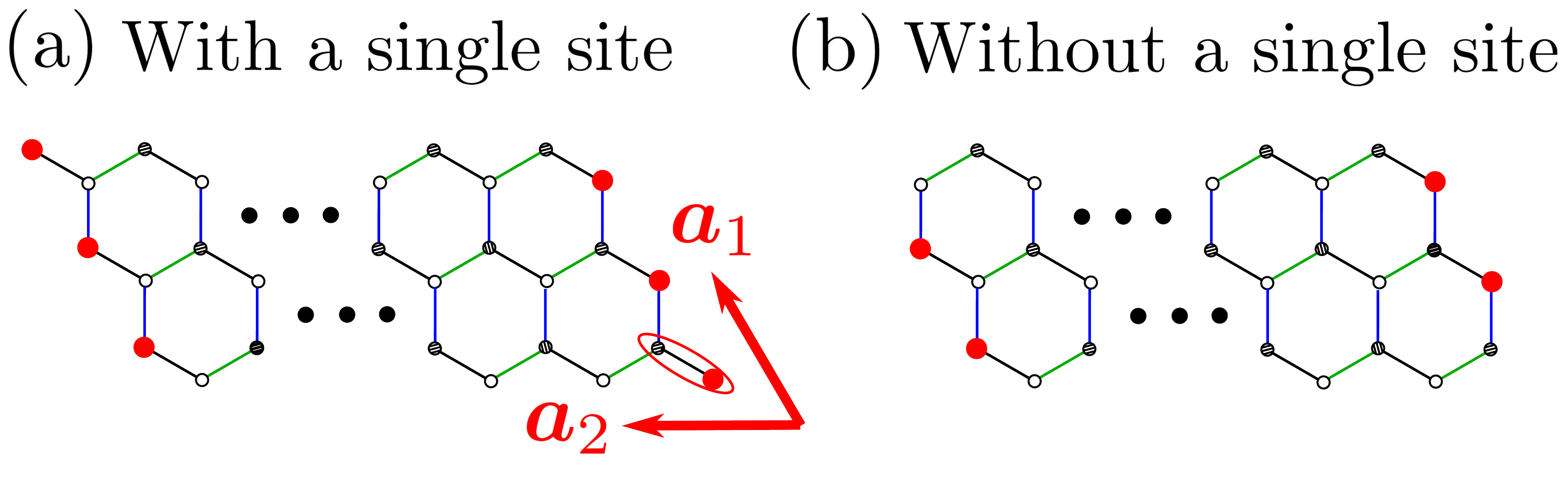}
    \caption{Schematic illustration of the honeycomb lattice under the full open boundary condition satisfying $N_{\bm{a}_2} > N_{\bm{a}_1}=3$. In the topological regime, Majorana zero modes (red) appear at both ends along the $\bm{a}_1$ direction. (a) When the edge is constructed such that the unit cells are preserved, six Majorana zero modes emerge. (b) When the edge termination is modified by removing a single atom at two corners, the number of Majorana zero modes is reduced by two, leaving four Majorana zero modes.}
    \label{Fig: main-concept}
    \end{figure}

For example, let us consider the quasi-1D systems illustrated in Fig.~\ref{Fig: main-concept}. In Fig.~\ref{Fig: main-concept} (a), the edge termination preserves the unit cell, so that there appear $2N_{\bm{a}_1}$ Majorana zero modes in total, i.e., $\omega=N_{\bm{a}_1}$ per edge. However, in Fig.~\ref{Fig: main-concept} (b), removing a single atom from both ends reduces the number of Majorana zero modes by two, resulting in $2N_{\bm{a}_1}-2$ Majorana zero modes. In this manner, the number of Majorana zero modes can be controlled through the choice of edge termination.

\section{nonlocal spin correlation}\label{sec: nonlocal_spin_corr}
In the topological regime, the quasi-1D system hosts Majorana zero modes localized at its ends. Their spatial localization pattern is sensitive to the relative strength of $J_x$ and $J_y$, while the number of Majorana zero modes varies depending on the choice of edge termination. In the following, we discuss spin correlations in the topological phase of the quasi-1D system and show that nonlocal spin correlations arise due to the presence of Majorana zero modes. To this end, we examine spin correlation functions defined by
\begin{align}
    \langle \text{GS} | S_i^{\mu} S_j^{\nu} | \text{GS} \rangle,
    \label{eq: two-point function}
\end{align}
where $| \text{GS} \rangle$ is a ground state wave function of the quasi-1D system. When the ground states have $g$-fold degeneracy, Eq.\ (\ref{eq: two-point function}) is replaced by the average value
\begin{align}
    \frac{1}{g}\sum_{\alpha=1}^{g} \langle \text{GS}_\alpha | S_i^{\mu} S_j^{\nu} | \text{GS}_\alpha \rangle, \label{eq:spin-correlation}
\end{align}
where $\langle \text{GS}_\alpha| \text{GS}_\beta \rangle=\delta_{\alpha\beta}$. This average value is invariant under unitary transformations in the ground-state sector. 

\subsection{Nonlocal spin correlation and fermion parity}
Here, we demonstrate that the presence of Majorana zero modes can lead to a nonlocal spin correlation in the quasi-1D systems. This correlation is naturally derived from the fermion parity of the Majorana zero modes and can therefore be interpreted as a spin analogue of Majorana-assisted nonlocal charge transfer \cite{Fu056402}.
We first discuss the case where only a single pair of Majorana zero modes exist at the left and right ends, i.e., the case where $\omega=\widetilde{W}_{\bm{a}_2}=1$. The extension to the situations with multiple Majorana zero modes will be discussed later. We note that, in addition to the pair of Majorana zero modes, there are additional localized Majorana fermions $\bar{\gamma}$ that lead to the $2^D$-fold degeneracy [Eq.~(\ref{Eq:extra-degeneracy})].

To connect spin correlations with the Majorana zero modes, we consider a nonlocal spin correlation between the $1$st and $N$th lattice sites, which belong to the B and A sublattices, respectively.
We use Eq.~(\ref{Spin-to-Majorana}) to write the product of the Majorana operators at the $1$st and $N$th sites as
\begin{align}
     \gamma_N \gamma_1 =\left[(\sigma^{x}_{N}) \prod^{N-1}_{j=1}(-\sigma^{z}_j)\right](-\sigma^{y}_{1}). 
     \label{eq:spin-corr1}
\end{align}
To proceed, we introduce the fermion parity operator for the Majorana zero modes, $P_{\text{MF}}$, and the total fermion parity operator, $P_{\text{total}}$, defined as
\begin{align}
    &P_{\text{MF}} \equiv  i\gamma_{\text{L}}\gamma_{\text{R}}, \label{eq:pmf}\\
    &P_{\text{total}} \equiv \prod_{j=1}^{N} (1 - 2n_j) = \prod_{j=1}^{N} (-\sigma^{z}_j), \label{eq:ptotal}
\end{align}
where $\gamma_{\text{L}}$ and $\gamma_{\text{R}}$ denote the Majorana zero modes localized at the left and right ends, respectively. These parity operators satisfy $P_{\rm MF}^2 =1$ and $P_{\rm total}^2=1$. The Majorana zero modes $\gamma_{\rm L}$ and $\gamma_{\rm R}$ are related to the Majorana operators $\gamma_1$ and $\gamma_N$ through the mode expansion in terms of the eigenstates of the Hamiltonian $H$ [Eq.~(\ref{Majorana-Kitaev})], 
\begin{subequations}
\label{eq:Mwf}
\begin{align}
&\gamma_{1} = u^{\text{R}}_1 \gamma_\text{R} + \text{(excited states)}, \\ 
&\gamma_{N} = u^{\text{L}}_N \gamma_\text{L} + \text{(excited states)},
\end{align}
\end{subequations}
where we separate the zero-energy from excited states, and $ u^{\text{R}}_1$ and $u^{\text{L}}_N$ are wave-function amplitudes of the zero modes. Substituting Eq.~(\ref{eq:Mwf}) into Eq.~(\ref{eq:spin-corr1}) and neglecting the contributions from excited states, we obtain 
\begin{align}
    u^{\text{L}}_N u^{\text{R}}_1  \gamma_\text{L} \gamma_\text{R} \approx \left[(\sigma^{x}_{N}) \prod^{N-1}_{j=1}(-\sigma^{z}_j)\right](-\sigma^{y}_{1}). 
     \label{eq:spin-corr2}
\end{align}
Hence, this relation can be rewritten as
\begin{align}
   -\sigma^{y}_{N} \sigma^{y}_{1} =
   u^{\text{L}}_N u^{\text{R}}_1 P_{\rm MF}P_{\rm total}
   + \ldots
   \label{eq:spin-corr3}
\end{align}
where the contributions from excited states are included in $\ldots$, and we have used the relation
\begin{align}
    \left[(\sigma^{x}_{N}) \prod^{N-1}_{j=1}(-\sigma^{z}_j)\right](-\sigma^{y}_{1}) = i \sigma_N^y \sigma_1^y P_{\rm total}.
\end{align}
Therefore, the product of spin operators $\sigma_N^y$ and $\sigma_1^y$ can be expressed directly in terms of the fermion parity operator. This relation [Eq.~(\ref{eq:spin-corr3})] constitutes one of the central results of this paper, indicating that the nonlocal correlation between Majorana zero modes give rise to the nonlocal spin correlation. Using this relation, the expectation value of the nonlocal spin correlation can be expressed as 
\begin{align}
     \langle \text{GS}_\alpha | S_N^{y} S_1^{y} | \text{GS}_\alpha \rangle \approx  -\frac{u^{\text{L}}_N u^{\text{R}}_1}{4} \langle \text{GS}_\alpha |  P_{\rm MF}P_{\rm total}| \text{GS}_\alpha \rangle,
      \label{eq:MF-spin}
\end{align}
where once again the excited-state contributions are neglected.

\subsection{Symmetry analysis}
\label{sec:symmetry_analysis}
Next, we discuss the symmetry constraints on $\langle \text{GS}_\alpha |  P_{\rm MF}P_{\rm total}| \text{GS}_\alpha \rangle $. We show that whether the nonlocal spin correlation is finite or vanishes depends on the ground-state degeneracy.
The following commutation relations are satisfied among the parity operators (\ref{eq:pmf}) and (\ref{eq:ptotal}), the Majorana zero modes (\ref{eq:Mwf}), the localized Majorana fermions $\bar{\gamma}_i$ that are not included in any of $W_p$ and lead to the $2^D$-fold degeneracy in Eq.~(\ref{Eq:extra-degeneracy}), the Hamiltonian (\ref{Majorana-Kitaev}), and the $\mathbb{Z}_2$ flux $W_p$:
\begin{align}
    &[P_{\rm MF},H]=[P_{\rm total},H] = [P_{\rm MF},P_{\rm total}]=0, \label{eq:commpp} \\
    &[P_{\rm MF},W_p]=[P_{\rm total},W_p] = [\bar{\gamma}_i,W_p]=0, \label{eq:commwp} \\
     &\{\gamma_{\rm L(R)},P_{\rm MF}\}=\{\gamma_{\rm L(R)},P_{\rm total}\} = [\gamma_{\rm L(R)},H]=0, \label{eq:commgr} \\
     &[\bar{\gamma}_i,P_{\rm MF}] =\{\bar{\gamma}_i,P_{\rm total}\} = [\bar{\gamma}_i,H]=0. \label{eq:commbargr} 
\end{align}
The commutation relation between $H$ and the other operators reflects the fermion-parity conservation and the zero-energy nature of $\gamma_{\rm L}$, $\gamma_{\rm R}$, and $\bar{\gamma}_i$. The commutation relations between $\gamma_{\rm L(R)}$, $P_{\rm MF}$, $P_{\rm total}$, $\bar{\gamma}_i$, and $W_p$ are derived from Eqs.~(\ref{eq:defwp}), (\ref{eq:pmf}), (\ref{eq:ptotal}), and (\ref{eq:Mwf}). 

Equation~(\ref{eq:commpp}) tells us that $H$, $P_{\rm MF}$, and $P_{\rm total}$ can be diagonalized simultaneously, and thus, the flux-free ground states are eigenstates of $P_{\rm MF}$ and $P_{\rm total}$.  Eq.~(\ref{eq:commwp}) implies that $P_{\rm MF}$, $P_{\rm total}$ and $\bar{\gamma}_i$ keep the flux configurations invariant. Finally, Eq.~(\ref{eq:commgr}) indicates that ground states possess at least a twofold degeneracy. To see this, we denote the ground state wave functions as $| \text{GS},p_{\rm MF}, p_{\rm total} \rangle $ with the eigenvalues $p_{\rm MF} =\pm 1$ and  $ p_{\rm total} = \pm 1$ of the parity operators $P_{\rm MF}$ and $P_{\rm total}$, respectively.
Then, the action of $\gamma_{\rm L}$ on a ground state leads to
\begin{align}
    \gamma_L| \text{GS},p_{\rm MF}, p_{\rm total} \rangle =| \text{GS},-p_{\rm MF}, -p_{\rm total} \rangle, 
\end{align}
up to a $U(1)$ phase factor, and the commutation relation $[\gamma_{L},H]=0$ implies that $|\text{GS}_1 \rangle \equiv| \text{GS},p_{\rm MF}, p_{\rm total} \rangle$ and $|\text{GS}_2 \rangle \equiv\gamma_L| \text{GS},p_{\rm MF}, p_{\rm total} \rangle$ have the same energy.
Furthermore, from Eq.~(\ref{eq:commbargr}), the existence of the decoupled (redundant) $\bar{\gamma}_i$ enforces at least fourfold degeneracy. This is because the action of $\bar{\gamma}_i$ on a ground state wave function yields
\begin{align}
    \bar{\gamma}_i | \text{GS},p_{\rm MF}, p_{\rm total} \rangle = | \text{GS},p_{\rm MF}, -p_{\rm total} \rangle,
\end{align}
up to a $U(1)$ phase factor. Thus, $|\text{GS}_1 \rangle$, $|\text{GS}_2 \rangle $, $|\text{GS}_3 \rangle \equiv \bar{\gamma_i}| \text{GS},p_{\rm MF}, p_{\rm total} \rangle $, and $|\text{GS}_4 \rangle \equiv \bar{\gamma_i} \gamma_L| \text{GS},p_{\rm MF}, p_{\rm total} \rangle $ are independent of each other and have the same energy.

When the ground-state degeneracy is only twofold due to $\gamma_{\rm L}$ and $\gamma_{\rm R}$, the nonlocal spin correlation retains a finite average value,  
\begin{align}
     \frac{1}{2}\sum_{\alpha=1}^{2} \langle \text{GS}_\alpha | S_N^{y} S_1^{y} | \text{GS}_\alpha \rangle
     &=\langle{\rm GS}_1|S_N^y S_1^y |{\rm GS}_1\rangle
     \nonumber\\
     &\approx  -\frac{p_{\rm MF} p_{\rm total} }{4} u^{\text{L}}_N u^{\text{R}}_1.
     \label{eq: nonlocal-spin-correlation}
\end{align}
This conclusion holds independently of the relative magnitudes of $J_x$ and $J_y$, as long as there are two Majorana zero modes $\gamma_{\rm L}$ and $\gamma_{\rm R}$. 
The parameters $J_x$ and $J_y$ enter through the coefficients $u^L_N$ and $u^R_1$, affecting only the quantitative amplitude of the average value.

The nonvanishing nonlocal spin correlation between the two opposite ends in the quasi-1D geometry in Eq.\ (\ref{eq: nonlocal-spin-correlation}) may look surprising in view of the fact that the excitations in the bulk are gapped.
This is no surprise, however, because the ground states are degenerate.
According to the general theorem stated in Ref.~\cite{Koma781}, the spin correlation function
\begin{equation}
    \frac{1}{g}\sum_{\alpha=1}^g
    \langle{\rm GS}_\alpha|S_N^y Q S_1^y
       |{\rm GS}_\alpha\rangle
    \label{eq: Koma-Hastings}
\end{equation}
must decay exponentially with the distance $N-1$ in gapped ground states,
where $Q$ is the projection operator onto excited states,
\begin{equation}
Q=1-\sum_{\alpha=1}^g
|{\rm GS}_\alpha\rangle \langle{\rm GS}_\alpha|.
\end{equation}
For the above twofold degenerate ground states $|{\rm GS}_1\rangle$ and $|{\rm GS}_2\rangle$ that are eigenstates of $P_{\rm total}$, Eq.\ (\ref{eq: Koma-Hastings}) implies
\begin{equation}
    \langle{\rm GS}_1|S_N^y S_1^y|{\rm GS}_1\rangle
    =\langle{\rm GS}_1|S_N^y|{\rm GS}_2\rangle
     \langle{\rm GS}_2|S_1^y|{\rm GS}_1\rangle,
     \label{eq: spin-correlation2}
\end{equation}
where we have used the fact that $S_N^y$ and $S_1^y$ change the total fermion parity with $\{P_{\text{total}},S^{y}_1\}=\{P_{\text{total}},S^{y}_N\}=0$, and therefore
\begin{equation}
    \langle{\rm GS}_\alpha|S_N^y|{\rm GS}_\alpha\rangle
    =\langle{\rm GS}_\alpha|S_1^y|{\rm GS}_\alpha\rangle
    =0.
\label{eq: spin-net}
\end{equation}
Note that one can show that $\langle S^{\mu}_j\rangle = 0$ for all $\mu$ and $j$ in our system, where $\langle \cdot \rangle$ implies the ground-state average value defined in Eq.~(\ref{eq:spin-correlation}) (see Appendix~\ref{app: symmetry-constrained-zero-magnetization}).

In contrast, if ground states are fourfold degenerate due to the presence of $\bar{\gamma}_i$ in addition to $\gamma_{\rm L}$ and $\gamma_{\rm R}$, the nonlocal spin correlation vanishes as
\begin{align}
     \frac{1}{4}\sum_{\alpha=1}^{4} \langle \text{GS}_\alpha | S_N^{y} S_1^{y} | \text{GS}_\alpha \rangle &\approx  \frac{p_{\rm MF} p_{\rm total} - p_{\rm MF} p_{\rm total}}{4 } u^{\text{L}}_N u^{\text{R}}_1\nonumber\\
     &=0.
\end{align}
 Consequently, the presence of the degeneracy arising from $\bar{\gamma}_i$ leads to the disappearance of the nonlocal spin correlation. In other words, to obtain a nonzero average value for the nonlocal spin correlation between $S_N^y$ and $S_1^y$, the extra degeneracy must be lifted.

The above discussion is extended to the case with multiple Majorana zero modes. In this case, the same relation as Eq.~(\ref{eq:MF-spin}) holds between the operators of Majorana zero modes located at the beginning and end of the Jordan--Wigner transformation sequence and the nonlocal spin correlation. Then, the additional degeneracy arising from the Majorana zero modes must be lifted so that ground states have twofold degeneracy associated with $\gamma_L$ and $\gamma_R$ and a nonzero average of the nonlocal spin correlation.

\subsection{Perturbation terms}

As discussed above, the symmetry constraints require that all redundant degeneracies be lifted for the nonlocal spin correlation to acquire a finite value. In this section, we demonstrate, using concrete lattice geometries, how such degeneracies can be removed. The redundant degeneracies originate from localized Majorana fermions or from multiple Majorana zero modes and depend on the lattice geometry. We denote a system with $N_{\bm{a}_1}$ sites along the $\bm{a}_1$ direction and $N_{\bm{a}_2}$ sites along the $\bm{a}_2$ direction as $(N_{\bm{a}_1},N_{\bm{a}_2})$. For illustration purposes, we consider the geometry shown in Fig.~\ref{Fig: main-concept} (b), where a single atom is removed from each end of the system. Thus, in the topological phase, $2N_{\bm{a}_1}-2$ Majorana zero modes appear.

In the following, we consider two lattice geometries $(2,N_{\bm{a}_2})$ and $(3,N_{\bm{a}_2})$ with $N_{\bm{a}_2} > 3$ and show that the $(3,N_{\bm{a}_2})$ geometry is a minimal setup to acquire the nonlocal spin correlation.

\subsubsection{\texorpdfstring{$(2,N_{\boldsymbol{a}_2})$ geometry}{(2, N_a2) geometry}}

\begin{figure}[t]
    \centering
    \includegraphics[scale=0.225]{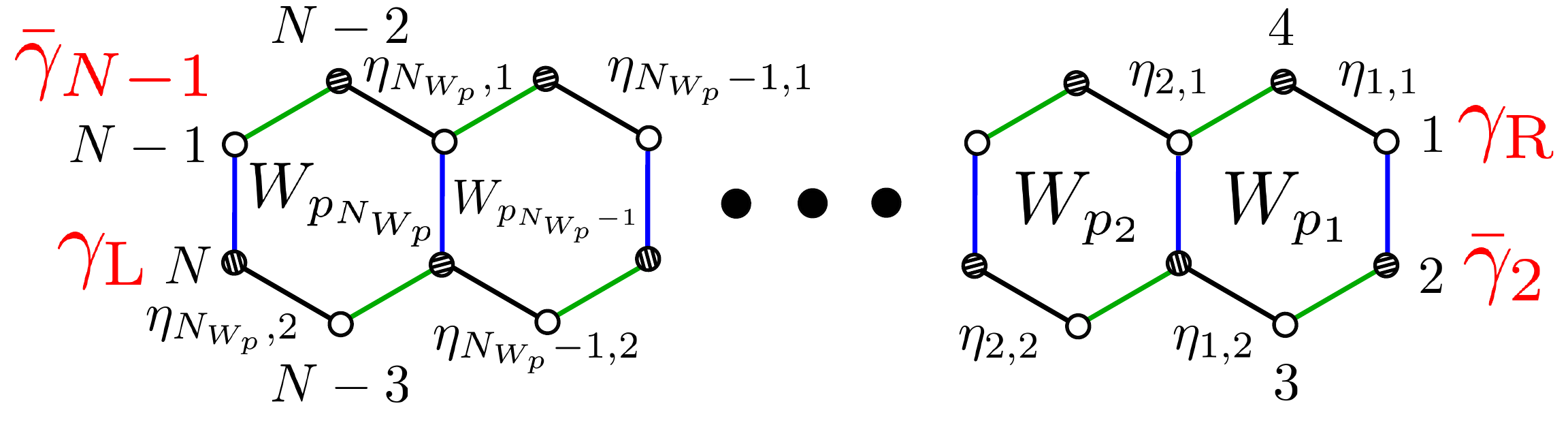}
    \caption{Configuration of quasi-1D Kitaev honeycomb model with the $(2,N_{\bm{a}_2})$ geometry. The numbers assigned to each site represent the labels used in the Jordan--Wigner transformation.  The number of plaquettes is $N_{W_p} = N_{\bm{a}_2}-1$, and the total number of sites is $N = 4N_{W_p}+2$.}
    \label{Fig: one-plaquette}
\end{figure}

We first consider the $(2,N_{\bm{a}_2})$ geometry shown in Fig.~\ref{Fig: one-plaquette}.
In this configuration, two Majorana zero modes, denoted by $\gamma_{\text{L}}$ and $\gamma_{\text{R}}$, appear at the opposite ends. We now evaluate the degeneracy arising from the localized Majorana fermions $\bar{\gamma}_i$. As seen in Fig.~\ref{Fig: one-plaquette}, the number of plaquettes is $N_{W_p} = N_{\bm{a}_2}-1$, and the total number of sites is $N = 4N_{W_p}+2$. Hence, from Eq.~(\ref{Eq:extra-degeneracy}), the degeneracy due to the localized Majorana fermions can be expressed as
\begin{align}
2^D = 2^{N_{W_p}+1}.
\label{additional-degeneracy}
\end{align}
The origin of these degeneracies can be understood as follows. The $\mathbb{Z}_2$ flux of the plaquettes is defined as $W_{p_j} = \eta_{j,1} \eta_{j,2}$ $(j=1,\cdots,N_{W_p})$. The flux-free condition requires $\eta_{j,1} \eta_{j,2}=1$ for all $j$. As shown in Fig.~\ref{Fig: one-plaquette}, neighboring plaquettes do not share $\eta_{j,k}$ ($k=1,2$) with each other, which allows independent choices of $\eta_{j,1} = \eta_{j,2} = \pm1$. Therefore, the $2^{N_{W_p}}$-fold degeneracy originates from these sign choices for individual plaquettes. In addition, the remaining twofold degeneracy arises from the two unpaired localized Majorana fermions at sites $2$ and $N-1$, denoted by $\bar{\gamma}_2$ and $\bar{\gamma}_{N-1}$ (see Fig.~\ref{Fig: one-plaquette}), which define the fermion-parity operator $i\bar{\gamma}_2\bar{\gamma}_{N-1}$.

From the above consideration, the $2^{N_{W_p}}$ degeneracy can be lifted by introducing local perturbations of the form
\begin{align}
V_{\eta} = \sum^{N_{W_p}}_{j=1} \lambda_j \eta_{j,1},
\label{additional-perturbation}
\end{align}
where $\lambda_j$ are constants. Each $\eta_{j,1}$ is expressed as a product of four spin operators from a plaquette via Eq.~(\ref{Spin-to-Majorana}). For example, $\eta_{1,1}$ is given by 
\begin{align}
\eta_{1,1} = i\bar{\gamma}_4\bar{\gamma}_1 = -\sigma^y_{4}\sigma^z_{3}\sigma^z_{2}\sigma^y_{1}, 
\end{align}
where the 1st and 3rd sites belong to the B sublattice, while the 2nd and 4th sites belong to the A sublattice. 
However, the twofold degeneracy due to the fermion parity operator $i\bar{\gamma}_2\bar{\gamma}_{N-1}$ cannot be removed by any local perturbations, since its lifting would require a nonlocal perturbation. Consequently, the nonlocal spin correlation vanishes in the $(2,N_{\bm{a}_2})$ geometry. 

\subsubsection{\texorpdfstring{$(3,N_{\boldsymbol{a}_2})$ geometry}{(3, N_a2) geometry}}
    
\begin{figure}[t]
    \centering
    \includegraphics[scale=0.24]{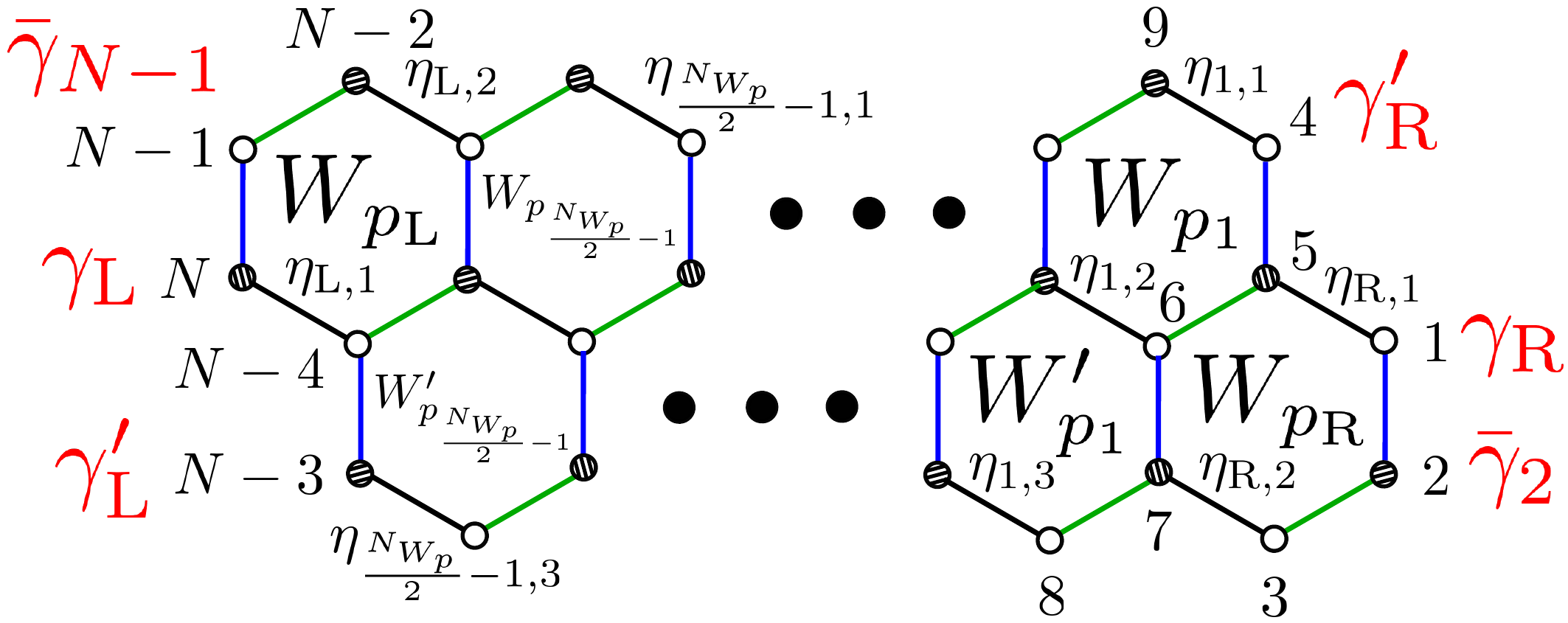}
    \caption{Configuration of quasi-1D Kitaev honeycomb model with the $(3,N_{\bm{a}_2})$ geometry. Each number shown on the sites corresponds to the label of the Jordan--Wigner transformation. The number of plaquettes is $N_{W_p} = 2N_{\bm{a}_2}-2$ and the total number of sites is $N = 3N_{W_p}+4$.}
    \label{Fig: two-plaquette}
\end{figure}

Next, we consider the $(3,N_{\bm{a}_2})$ geometry as shown in Fig.~\ref{Fig: two-plaquette}.
In this case, four Majorana zero modes appear at the boundaries, denoted by $\gamma_L$, $\gamma_L'$, $\gamma_R$, $\gamma_R'$  (see Fig.~\ref{Fig: two-plaquette}). Here, $\gamma_L'$ and $\gamma_R'$ represent additional Majorana zero modes that must be lifted by local perturbations in order to induce a finite nonlocal spin correlation. 
In this configuration, the degeneracy associated with the localized Majorana fermions is given by 
\begin{align}
2^D = 2^{\frac{N_{W_p}}{2}+2},
\label{additional-degeneracy2}
\end{align}
where the number of plaquettes is $N_{W_p} = 2N_{\bm{a}_2}-2$ and the total number of sites is $N = 3N_{W_p}+4$. Each plaquette is labeled as illustrated in Fig.~\ref{Fig: two-plaquette}: the plaquettes at the left and right edges are denoted by $W_{p_{\text{L}}} = \eta_{\text{L},1} \eta_{\text{L},2}$ and $W_{p_{\text{R}}}=\eta_{\text{R},1} \eta_{\text{R},2}$, respectively, while the remaining plaquettes are defined as $W_{p_j} = \eta_{j,1}\eta_{j,2}$ and $W_{p_j}'=\eta_{j,2}\eta_{j,3}$ ($j=1,\cdots, \frac{N_{W_p}}{2}-1$).

The origin of these degeneracies can be understood as follows. As shown in Fig.~\ref{Fig: two-plaquette}, neighboring plaquettes $W_{p_j}$ and $W_{p_j}'$ share the bond variable $\eta_{j,2}$, which allows independent choices of $\eta_{j,1} = \eta_{j,2} = \eta_{j,3} = \pm 1$ under the flux-free constraints $W_{p_j} = W_{p_j}'=+1$. This results in a $2^{(N_{W_p}-2)/2}$ degeneracy. In contrast, the edge plaquettes $W_{p_{\text{L}}}$ and $W_{p_{\text{R}}}$ do not share any $\eta$ bonds with other plaquettes, permitting independent choices of $\eta_{\text{L},1} = \eta_{\text{L},2} = \pm 1$ and $\eta_{\text{R},1} = \eta_{\text{R},2} = \pm 1$, which yield an additional fourfold degeneracy. Furthermore, a twofold degeneracy arises from the two unpaired localized Majorana fermions at sites $2$ and $N-1$, denoted by $\bar{\gamma}_2$ and $\bar{\gamma}_{N-1}$, defining the fermion parity operator $i\bar{\gamma}_2\bar{\gamma}_{N-1}$. Combining these contributions, we obtain 
\begin{align}
    D=\frac{N_{W_p}}{2}-1+2+1=\frac{N_{W_p}}{2}+2.
\end{align}

We now discuss local perturbations that lift these degeneracies. 
Analogous to the $(2,N_{\bm{a}_2})$ geometry, the $2^{(N_{W_p}-2)/2}$ degeneracy can be removed by introducing local perturbations of the form
\begin{align}
V_1 = \sum^{\frac{N_{W_p}}{2}-1}_{j=1} \lambda_j \eta_{j,1},
\label{additional-perturbation2}
\end{align}
where $\lambda_j$ are constants. Using Eq.~(\ref{Spin-to-Majorana}), each $\eta_{j,1}$ is expressed as a product of six spin operators. For instance,
\begin{align}
\eta_{1,1} = i\bar{\gamma}_9\bar{\gamma}_4 =-\sigma^y_{9}\sigma^z_{8}\sigma^z_{7}\sigma^z_{6}\sigma^z_{5}\sigma^y_{4}, 
\end{align}
where the 4th, 6th, and 8th sites belong to the B sublattice, while the 5th, 7th, and 9th sites belong to the A sublattice.

The fourfold degeneracy associated with the edge plaquettes $W_{p_L}$ and $W_{p_R}$ is lifted by adding local perturbations
\begin{align}
V_2 =  \lambda_{\text{L}} \eta_{\text{L},1} + \lambda_{\text{R}} \eta_{\text{R},1},
\label{additional-perturbation3}
\end{align}
where $\eta_{\text{L},1}$ and $\eta_{\text{R},1}$ are given by
\begin{align}
&\eta_{\text{L},1} = i\bar{\gamma}_N\bar{\gamma}_{N-4} = \sigma^y_{N}\sigma^z_{N-1}\sigma^z_{N-2}\sigma^z_{N-3}\sigma^y_{N-4}, \label{eq:v2-spin1}\\ 
&\eta_{\text{R},1} = i\bar{\gamma}_5\bar{\gamma}_1 = \sigma^y_{5}\sigma^z_{4}\sigma^z_{3}\sigma^z_{2}\sigma^y_{1}, \label{eq:v2-spin2}
\end{align}
Here, the 1st, 3rd, 4th, $(N-4)$th, and $(N-1)$th sites belong to the $B$ sublattice, while the 2nd and 5th, $(N-3)$th, $(N-2)$th, and $N$th sites belong to the $A$ sublattice. 

Finally, the twofold degeneracy originating from the unpaired localized Majorana fermions is lifted by introducing local perturbations between $\gamma_{\text{L}}' $ and $\bar{\gamma}_{N-1} $, and between $\gamma_{\text{R}}' $  and $\bar{\gamma}_2$, 
\begin{align}
V_3 &= i\rho_\text{L}\gamma_{N-3}\bar{\gamma}_{N-1} + i\rho_\text{R} \gamma_{4}\bar{\gamma}_{2}, 
\label{additional-perturbation4}
\end{align}
since  $\gamma_{\text{L}}' $ and $\gamma_{\text{R}}' $ are localized near $(N-3)$th and 4th sites, respectively.
Here, $\rho_\text{L}$ and $\rho_\text{R}$ are constants. In terms of the spin operators, these perturbations are expressed as
\begin{align}
&i\gamma_{N-3}\bar{\gamma}_{N-1} = \sigma^x_{N-1} \sigma^z_{N-2} \sigma^y_{N-3}, \label{eq:v3-spin1}\\
&i\gamma_{4}\bar{\gamma}_{2} = \sigma^y_4 \sigma^z_3 \sigma^x_2. \label{eq:v3-spin2}
\end{align}
One can verify that $[V_3,\gamma_{\text{L}}]=[V_3,\gamma_{\text{R}}]=0$ for $N \rightarrow \infty$.
Equations (\ref{eq:v2-spin1}), (\ref{eq:v2-spin2}), (\ref{eq:v3-spin1}), and (\ref{eq:v3-spin2}) break time-reversal symmetry due to the presence of five-spin and three-spin products. Nevertheless, the local magnetization of the ground states remains zero even when these terms are included (See Sec. \ref{sec: numerics} for a numerical calculation and Appendix \ref{app: symmetry-constrained-zero-magnetization} for a symmetry argument).

All redundant degeneracies are thus lifted by introducing the local perturbations $V_1$, $V_2$, and $V_3$, leaving only the twofold degeneracy associated with $\gamma_L$ and $\gamma_R$. Consequently, the nonlocal spin correlation is expected to become finite. In the next section, we will demonstrate this finite nonlocal spin correlation numerically. 

For a general lattice geometry $(N_{\bm{a}_1},N_{\bm{a}_2})$ satisfying $N_{\bm{a}_1} < N_{\bm{a}_2}$, the nonlocal spin correlation remains finite when $N_{\bm{a}_1}$ is an odd integer, after local perturbations leave only a twofold degeneracy as in the $(3,N_{\bm{a}_2})$ geometry. 

We also comment on the physical realization of such local perturbations. In candidate Kitaev materials, non-Kitaev interactions, such as Heisenberg and symmetric off-diagonal exchange interactions, are generally present, while externally applied electric and magnetic fields, $\bm{E}$ and $\bm{B}$, can further induce additional interaction terms \cite{Chari134444, Furuya013228}. Within the flux-free sector, these ingredients generate effective couplings among multiple spin operators in perturbation theory: six-spin terms arise at cubic order in the electric field ($E^3$), five-spin terms at the mixed order ($E^2B$), and three-spin terms emerge at both the mixed order $EB$ and cubic order in the magnetic field ($B^3$). Therefore, the local perturbations considered in our analysis are, in principle, realizable by combining these non-Kitaev couplings with controlled electric and magnetic fields.

\subsection{Numerical calculation}\label{sec: numerics}

\begin{table}[t]
	\caption{The parameters used in the exact diagonalization, where we set $\lambda' \equiv \lambda_1 = \lambda_2$ in Eq.~(\ref{additional-perturbation2}).}
	\begin{tabular}{ccccccccc}
		\hline\hline
		$J_x$ & $J_y$ & $J_z$ & $\lambda$ & $\lambda'$ & $\lambda_{\text{L}}$ & $\lambda_{\text{R}}$ & $\rho_{\text{L}}$ & $\rho_{\text{R}}$ 
		\\
		\hline
  	$1.6$ & $1.39$ & $0.01$ & $-0.01$ & $0.01$ & $0.01$ & $0.01$ & $0.001$ & $0.001$ 
        \\
		\hline \hline
	\end{tabular}
	\label{Tab: parameters}
\end{table}

We consider the Hamiltonian (\ref{Kitaev}) with the local perturbation terms (\ref{additional-perturbation2}), (\ref{additional-perturbation3}), and (\ref{additional-perturbation4})
\begin{align}
    H + V_1 + V_2 + V_3. \label{eq:hami_tot}
\end{align}
We demonstrate the nonlocal spin correlation in quasi-1D Kitaev honeycomb model with the $(3,N_{\bm{a}_2})$ lattice geometry by numerically diagonalizing Eq.~(\ref{eq:hami_tot}). The total number of sites is set to $N=22$, corresponding to $N_{\bm{a}_1}=3$ and $N_{\bm{a}_2}=4$, as illustrated in Fig.~\ref{Fig: two-plaquette}. 
Here, due to the computational cost, the system size accessible to exact diagonalization is limited. As a result, the overlap between the Majorana zero modes at the left and right ends cannot be neglected, leading to a small finite energy splitting. As described in the Appendix \ref{app:detail}, when $N_{\bm{a}_2} < 5$, the finite-size-induced energy splitting becomes larger than the minimal flux excitation energy. To overcome this problem, we introduce a perturbation term of the form
\begin{align}
V = \lambda\left[\sum^{2}_{i=1}(W_{p_i}+W'_{p_i}) + W_{p_{\text{L}}}+W_{p_{\text{R}}}\right],
\label{eq:perturbation_flux}
\end{align}
which increases the flux excitation energy. When $\lambda <0$ and $|\lambda|$ exceeds the energy splitting, this perturbation stabilizes the flux-free sector, yielding results that are effectively equivalent to those for $N_{\bm{a}_2} \ge 5$.
In the following, the model parameters are chosen as listed in Table~\ref{Tab: parameters}.

The correspondence between the ground-state degeneracy obtained from exact diagonalization and the perturbation terms is summarized in Table~\ref{Tab: groundstate_deg}. As expected, the ground-state degeneracy becomes two when all perturbation terms are included. 
\begin{table}[t]
	\caption{The ground-state degeneracy under the local perturbations defined in Eqs.~(\ref{additional-perturbation2}), (\ref{additional-perturbation3}), (\ref{additional-perturbation4}), and (\ref{eq:perturbation_flux}).}
	\begin{tabular}{ccccc}
		\hline\hline
   	$ \ V \ $ & $ \ V+V_1\ $ & $ \ V+V_1+V_2\ $ & $V+V_1+V_2+V_3$
		\\
		\hline
  	$128$ & $32$ & $8$ & $2$ 
        \\
		\hline \hline
	\end{tabular}
	\label{Tab: groundstate_deg}
\end{table}

Using this twofold‑degenerate ground state $|\text{GS}_1 \rangle$ and $|\text{GS}_2 \rangle$, the expectation value of $P_{\text{total}}$, $\sigma_1^y$, $\sigma_N^y$, and their correlation are calculated as follows.
\begin{subequations}
\begin{align}
&\langle \text{GS}_1|P_{\text{total}}|\text{GS}_1\rangle = -1, \quad \langle \text{GS}_2|P_{\text{total}}|\text{GS}_2\rangle = 1, \\ 
&\langle \text{GS}_1|\sigma^y_1|\text{GS}_2\rangle \approx 0.157 + 0.637i,
\label{results1}\\ 
&\langle \text{GS}_{1}|\sigma^y_1|\text{GS}_{1}\rangle = \langle \text{GS}_{2}|\sigma^y_1|\text{GS}_{2}\rangle = 0, \\ 
&\langle \text{GS}_2|\sigma^y_N|\text{GS}_1\rangle \approx 0.157 - 0.637i,
\label{results2}
\\ 
&\langle \text{GS}_{1}|\sigma^y_N|\text{GS}_{1}\rangle = \langle \text{GS}_{2}|\sigma^y_N|\text{GS}_{2}\rangle =0, \\ \
&\langle \text{GS}_1|\sigma^y_1\sigma^y_N|\text{GS}_1\rangle = \langle \text{GS}_2|\sigma^y_1\sigma^y_N|\text{GS}_2\rangle \approx 0.43,
\label{results3}
\end{align}
\end{subequations}
where the first nontrivial two digits are written in Eqs.~(\ref{results1}), (\ref{results2}), and (\ref{results3}), and "$=$" means that equalities hold up to numerical errors of order $10^{-7}$. 
Thus, the average of expectation values is obtained as
\begin{align}
    &\frac{1}{2} \sum_{\alpha=1}^2 \langle \text{GS}_\alpha|\sigma^y_1|\text{GS}_\alpha\rangle = 0,\nonumber \\
    &\frac{1}{2} \sum_{\alpha=1}^2 \langle \text{GS}_\alpha|\sigma^y_N|\text{GS}_\alpha\rangle = 0, \nonumber \\
    &\frac{1}{2} \sum_{\alpha=1}^2 \langle \text{GS}_\alpha|\sigma^y_1\sigma^y_N|\text{GS}_\alpha\rangle \approx 0.43, 
    \label{eq: main-results}
\end{align}
indicating that the nonlocal spin correlation emerges in the absence of net local magnetization. These results are consistent with Eqs.~(\ref{eq: nonlocal-spin-correlation}), (\ref{eq: spin-correlation2}), and (\ref{eq: spin-net}).

Finally, we confirm the emergence of nonlocal spin correlation by introducing local magnetic fields. A local magnetic field is applied at the right edge,
\begin{align}
V_B = \pm |B_{\text{R}}| S^y_1,
\label{Zeeman}
\end{align}
to lift the remaining twofold degeneracy. Since $S^y_1 \propto \gamma_{\text{R}}$, the fermion parity $P_{\text{MF}}=i\gamma_{\text{L}}\gamma_{\text{R}}$ is no longer conserved. By numerically diagonalizing the  Hamiltonian
\begin{align}
   H + V_1 + V_2 + V_3 + V + V_{B}, 
\end{align}
we obtain a single ground state $|\text{GS}_+\rangle$ or $|\text{GS}_-\rangle$, where we put $|B_{\text{R}}|=0.0002$, and the subscript $\pm$ denotes the sign of the Zeeman coupling in Eq.~(\ref{Zeeman}). The corresponding expectation values of the spin operator at the right and left edges are
\begin{align}
\langle\text{GS}_{\pm}| \sigma^y_1 |\text{GS}_{\pm}\rangle = \langle\text{GS}_{\pm}| \sigma^y_N |\text{GS}_{\pm}\rangle \approx \mp 0.656,
\label{numerical-results2}
\end{align}
where the value $0.656 \approx \sqrt{0.43}$ corresponds to the Majorana wave functions $u^{\text{L}}_N$ and $u^{\text{R}}_1$ as shown in Fig.~\ref{Fig: wavefunction}.

\section{Conclusion}
We have revealed the Majorana-assisted nonlocal spin correlation in quasi-1D Kitaev spin liquids arising from edge-localized Majorana zero modes. In our explicit construction, the Majorana fermion parity and total fermion parity operators are mapped to the product of spin operators. The former is related to a nonlocal spin string operator, and the latter is the product of spin operators over all sites. As a result, the combination of these fermion parity operators gives the nonlocal spin correlation. 
In addition to this general expression of the nonlocal spin correlation, we identified the conditions under which such nonlocal spin correlations are finite in the ground states by lifting unwanted degeneracies with local perturbations while preserving the flux‑free constraint. We have shown that the nonlocal spin correlation emerges in the flux-free ground states due to total fermion parity conservation and topologically protected Majorana zero modes, establishing nonlocal spin correlation as an analog of Majorana‑assisted nonlocal charge transfer.

We also frame the edge physics within a finite-size topology framework that complements the 1D topology of the quasi-1D Kitaev honeycomb model. Projecting to the low-energy sector yields an effective class-BDI Hamiltonian that reproduces the phase boundaries of the flux-free states. From this perspective, Majorana zero modes appear as edge states of hybridized Majorana flat-band states, and are stabilized by finite‑size-induced topology. The spectrum then exhibits a continuous crossover from a finite-size topological regime (hybridized Majorana flat bands inside the bulk gap) to the intrinsic 1D regime (only zero modes inside the bulk gap). 

\section{Acknowledgments}
Y. Y. thanks Masahiro O. Takahashi, Takahiro Morimoto, Joji Nasu, and Hiroaki Matsueda for fruitful discussions. Y. Y. also thanks Tsutomu Momoi and Hitoshi Seo for their helpful advice in numerical calculations of the spin system.
This work was supported by JSPS KAKENHI (Grants No.\ JP19K03680, No.\ JP22K03478, No.\ 24K00557, No.\ JP25K07161, No.\ JP25K23353, and No.\ JP24H02231) and JST CREST (Grant No.\ JPMJCR19T2). The numerical calculations in the spin model have been performed using the HOKUSAI supercomputer system in RIKEN.

\appendix
 
\section{Examples of the labeling used in the Jordan--Wigner transformation}
\label{app: Jordan--Wigner}

\begin{figure}[t]
    \centering
    \includegraphics[scale=0.315]{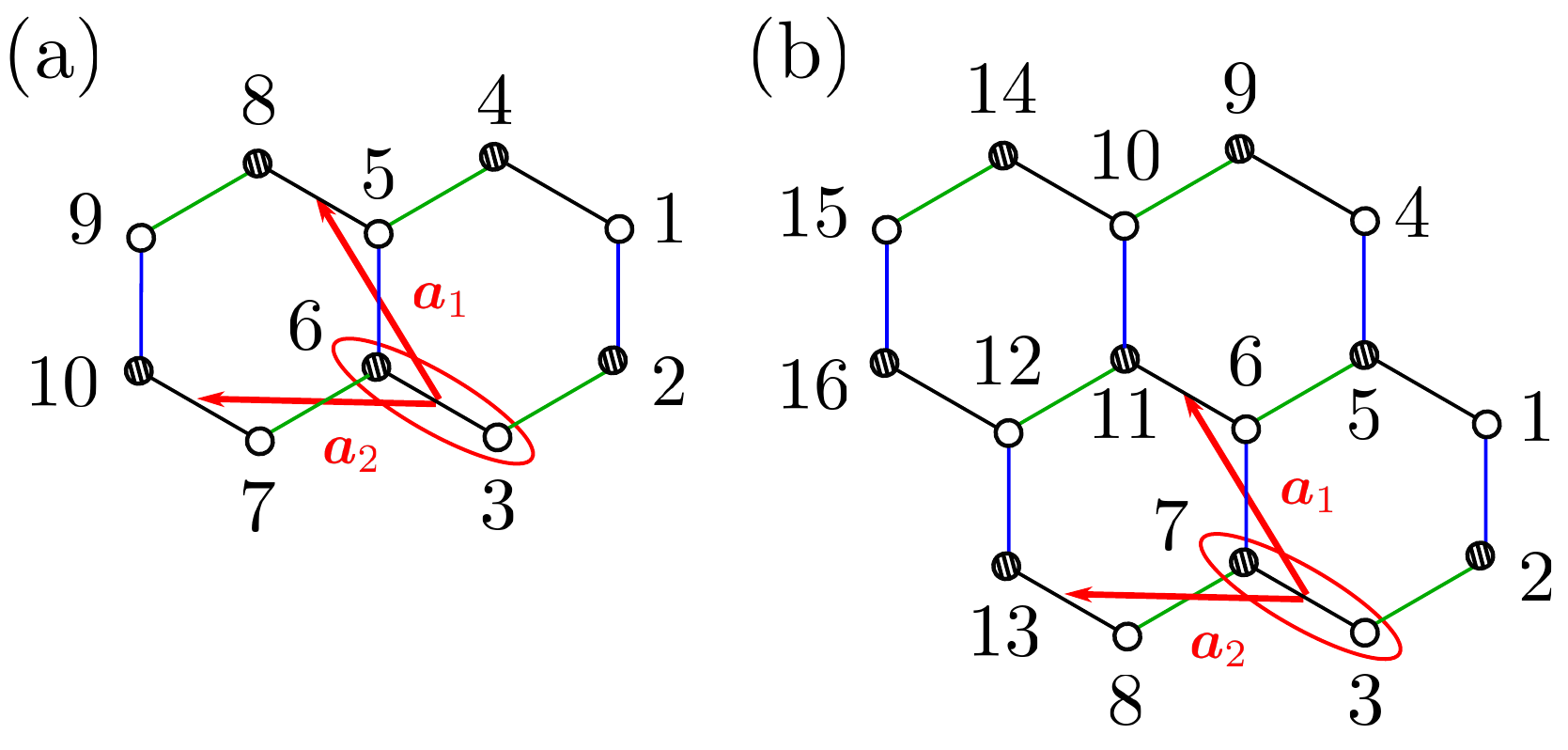}
    \caption{Examples of the labeling used in the Jordan--Wigner transformation for the Kitaev honeycomb model with open boundary conditions. The number denotes the labeling of each lattice site.}
    \label{fig: Jordan--Wigner}
\end{figure}

Here, we show examples of the labeling used in the Jordan--Wigner transformation for the Kitaev honeycomb model with open boundary conditions.
The labeling in the Jordan--Wigner transformation depends on the lattice geometry. In Fig.~\ref{fig: Jordan--Wigner} (a), the labeling corresponds to the case with one plaquette along $\bm{a}_1$ direction and two plaquettes along the $\bm{a}_2$ direction, while in Fig.~\ref{fig: Jordan--Wigner} (b), it corresponds to the case with two plaquettes along both the $\bm{a}_1$ and $\bm{a}_2$ directions. 

\section{Derivation of Eq.~(\ref{Sizeeffect-effective-Hamiltonian})}\label{app: finite-size}
Here, we show the derivation of Eq.~(\ref{Sizeeffect-effective-Hamiltonian}). We first examine the exact solution of the Majorana flat band in Eq.~(\ref{Kitaev-Hamiltonian}) for the $A_x$ phase ($|J_x| > |J_y|+|J_z|$). The argument is based on Refs. \cite{Thakurathi235434, Mizoguchi184418}. We consider the system with the open boundary condition only in the $\bm{a}_1$ direction. The Hamiltonian is given by Eq.~(\ref{tight-binding}). We search for a zero energy solution which satisfies
\begin{align}
H(k_{\bm{a}_2})\left(
\begin{array}{c}
u_{1,A}(k_{\bm{a}_2}) \\
u_{1,B}(k_{\bm{a}_2}) \\
u_{2,A}(k_{\bm{a}_2}) \\
u_{2,B}(k_{\bm{a}_2}) \\
\vdots \\
u_{N_{\bm{a}_1},A}(k_{\bm{a}_2}) \\
u_{N_{\bm{a}_1},B}(k_{\bm{a}_2})
\end{array}
\right) = 0,
\label{zero-energy}
\end{align}
which leads to the following equations:
\begin{align}
(J_y e^{ik_{\bm{a}_2}}+J_z) u_{r_{\bm{a}_1},B}(k_{\bm{a}_2}) + J_x u_{r_{\bm{a}_1}+1,B}(k_{\bm{a}_2})&=0, \\ 
J_x u_{r_{\bm{a}_1},A}(k_{\bm{a}_2}) + (J_y e^{-ik_{\bm{a}_2}}+J_z) u_{r_{\bm{a}_1}+1,A}(k_{\bm{a}_2})&=0, 
\end{align}
where $1 \le r_{\bm{a}_1} \le N_{\bm{a}_1}-1$ and we impose the boundary conditions such that $u_{1,A} = u_{N_{\bm{a}_1},B} = 0$. 
In the $N_{\bm{a}_1} \rightarrow \infty$ limit, we find the solutions as \cite{Thakurathi235434, Mizoguchi184418}
\begin{align}
u_{r_{\bm{a}_1},B}(k_{\bm{a}_2}) &= X(k_{\bm{a}_2})[\zeta(k_{\bm{a}_2})]^{r_{\bm{a}_1}-1}, \label{eigenstate-flatbandA} \\
u_{r_{\bm{a}_1},A}(k_{\bm{a}_2}) &= X(k_{\bm{a}_2}) [\zeta(-k_{\bm{a}_2})]^{N_{\bm{a}_1}-1-(r_{\bm{a}_1}-1)},
\label{eigenstate-flatbandB}
\end{align}
where $X$ and $\zeta$ are given by Eqs.~(\ref{eq:Xfn}) and (\ref{eq:zetafn}). The normalization factor $X$ is chosen in such a way that Eqs.~(\ref{eigenstate-flatbandA}) and (\ref{eigenstate-flatbandB}) satisfy $\sum^{N_{\bm{a}_1}}_{r_{\bm{a}_1}=1}|u_{r_{\bm{a}_1},s}(k_{\bm{a}_2})|^2=1$ $(s=A,B)$.
The zero energy state wave function can be written in the following form:
\begin{align}
\bm{u}_{A}(k_{\bm{a}_2}) &\equiv [u_{1,A}(k_{\bm{a}_2}), 0, u_{2,A}(k_{\bm{a}_2}), 0, \cdots , u_{N_{\bm{a}_1},A}(k_{\bm{a}_2}), 0]^{\text{T}},  \\
\bm{u}_{B}(k_{\bm{a}_2}) &\equiv [0, u_{1,B}(k_{\bm{a}_2}), 0, u_{2,B}(k_{\bm{a}_2}), \cdots , 0, u_{N_{\bm{a}_1},B}(k_{\bm{a}_2})]^{\text{T}},
\end{align}
where each vector satisfies
\begin{align}
\Gamma\bm{u}_{A}(k_{\bm{a}_2}) &= \bm{u}_{A}(k_{\bm{a}_2}), \\
\Gamma\bm{u}_{B}(k_{\bm{a}_2}) &= -\bm{u}_{B}(k_{\bm{a}_2}).
\end{align}
with $\Gamma=\text{diag}[1,-1,...,1,-1]$ being the chiral operator.

Using the wave function of the Majorana flat bands, the effective Hamiltonian (\ref{Sizeeffect-effective-Hamiltonian}) is derived as
\begin{align}
\tilde{H}(k_{\bm{a}_2}) &= \left[
\begin{array}{cc}
\bm{u}^{\dagger}_{A}(k_{\bm{a}_2})H(k_{\bm{a}_2})\bm{u}_{A}(k_{\bm{a}_2}) & \bm{u}^{\dagger}_{A}(k_{\bm{a}_2})H(k_{\bm{a}_2})\bm{u}_{B}(k_{\bm{a}_2})\\
\bm{u}^{\dagger}_{B}(k_{\bm{a}_2})H(k_{\bm{a}_2})\bm{u}_{A}(k_{\bm{a}_2})  & \bm{u}^{\dagger}_{B}(k_{\bm{a}_2})H(k_{\bm{a}_2})\bm{u}_{B}(k_{\bm{a}_2})
\end{array}
\right] ,
\end{align}
where each component is calculated as
\begin{widetext}
\begin{align}
\bm{u}^{\dagger}_{s}(k_{\bm{a}_2})H(k_{\bm{a}_2})\bm{u}_{s}(k_{\bm{a}_2}) &=0, \quad (s=A,B), \\ \nonumber
\bm{u}^{\dagger}_{A}(k_{\bm{a}_2})H(k_{\bm{a}_2})\bm{u}_{B}(k_{\bm{a}_2}) &= -\frac{iJ_x}{2}\sum^{N_{\bm{a}_1}-1}_{r_{\bm{a}_1}=1}u^{*}_{r_{\bm{a}_1}-1,A}(k_{\bm{a}_2})u_{r_{\bm{a}_1},B}(k_{\bm{a}_2}) -\frac{i(J_ye^{ik_{\bm{a}_2}}+J_z)}{2}\sum^{N_{\bm{a}_1}}_{r_{\bm{a}_1}=1}u^{*}_{r_{\bm{a}_1},A}(k_{\bm{a}_2})u_{r_{\bm{a}_1},B}(k_{\bm{a}_2}) \\ \nonumber
&=(X(k_{\bm{a}_2}))^2 \Bigg[-\frac{iJ_x}{2}\sum^{N_{\bm{a}_1}-1}_{r_{\bm{a}_1}=1} \left(-\frac{J_y e^{ik_{\bm{a}_2}}+J_z}{J_x}\right)^{N_{\bm{a}_1}-1-(r_{\bm{a}_1}-2)}\left(-\frac{J_y e^{ik_{\bm{a}_2}}+J_z}{J_x}\right)^{r_{\bm{a}_1}-1} \\ \nonumber
&\quad -\frac{i(J_y e^{ik_{\bm{a}_2}}+J_z)}{2}\sum^{N_{\bm{a}_1}}_{r_{\bm{a}_1}=1} \left(-\frac{J_y e^{ik_{\bm{a}_2}}+J_z}{J_x}\right)^{N_{\bm{a}_1}-1-(r_{\bm{a}_1}-1)}\left(-\frac{J_y e^{ik_{\bm{a}_2}}+J_z}{J_x}\right)^{r_{\bm{a}_1}-1}\Bigg] \\ \nonumber
&= (X(k_{\bm{a}_2}))^2 \Bigg[-\frac{iJ_x}{2}(N_{\bm{a}_1}-1) \left(-\frac{J_y e^{ik_{\bm{a}_2}}+J_z}{J_x}\right)^{N_{\bm{a}_1}} + \frac{iJ_x}{2}N_{\bm{a}_1} \left(-\frac{J_y e^{ik_{\bm{a}_2}}+J_z}{J_x}\right)^{N_{\bm{a}_1}}\Bigg] \\ \nonumber
&= \frac{iJ_x}{2}(X(k_{\bm{a}_2}))^2 \Bigg[ \left(-\frac{J_y e^{ik_{\bm{a}_2}}+J_z}{J_x}\right)^{N_{\bm{a}_1}} \Bigg] \\
&=\frac{i J_x}{2}(X(k_{\bm{a}_2}))^2 \left(\zeta(k_{\bm{a}_2}) \right)^{N_{\bm{a}_1}}.
\end{align}
\end{widetext}

\section{Symmetry constraints on magnetization}\label{app: symmetry-constrained-zero-magnetization}
In this appendix, we demonstrate that the local magnetization vanishes for all spin components and at all lattice sites in the ground states, provided that the following three conditions are satisfied: (i) the system is in the flux-free sector; (ii) only two Majorana zero modes, $\gamma_{\text{L}} $ and $ \gamma_{\text{R}}$, are present, leading to a twofold-degenerate ground state denoted by $|\text{GS}_1\rangle$ and $|\text{GS}_2\rangle \equiv \gamma_{\rm L} |\text{GS}_1\rangle$, as discussed in Sec. \ref{sec:symmetry_analysis}; (iii) a finite energy gap exists between the ground states and first excited states. We show below that $\langle S^x_j \rangle = \langle S^y_j \rangle = 0$ under conditions (i) and (ii), while $\langle S^z_j \rangle = 0$ is approximately satisfied when conditions (i)-(iii) are satisfied. Here, the average value is defined by $\langle S^\mu_j \rangle = \frac{1}{2}\sum_{\alpha=1}^2  \langle \text{GS}_\alpha | S^\mu_j |\text{GS}_\alpha \rangle$.

First, we consider the average value of $S^x_j$ and $S^y_j$ for all sites $j$, which include an odd number of Majorana operators as in Eq.~(\ref{Spin-to-Majorana}). Since $\{\gamma_{a}, S_j^x\} =\{\gamma_{a}, S_j^y\} =0 \text{ for } a=\text{L or R}$, we obtain
\begin{align}
&\sum_{\alpha=1}^{2} \langle \text{GS}_\alpha | S^x_{j}  | \text{GS}_\alpha \rangle =\sum_{\alpha=1}^{2} \langle \text{GS}_\alpha | S^y_{j}  | \text{GS}_\alpha \rangle = 0,
\end{align}
where $|\text{GS}_2\rangle = \gamma_{\text{L}}| \text{GS}_1\rangle$ for $j = 1,2,...,N$.

Next, we consider the average value of $S^z_j = i s_j \gamma_j \bar{\gamma}_j$ with $s_j$  taking $+1$ ($-1$) when $j$ is an A (B) site.  The $S^z_j$ operators are classified into whether they flip the sign of the $\mathbb{Z}_2$ flux ($\{S^z_j, W_{p}\}=0$) or not. From Eq.~(\ref{Spin-to-Majorana}), $S^z_j$ is the flux-flipping operator when $\bar{\gamma}_j$ is included in one of the $\mathbb{Z}_2$ fluxes. Since the one-flux excitation states are orthogonal to the flux-free state, it is satisfied that
\begin{align}
\langle \text{GS}_\alpha| S^z_{j} |\text{GS}_\alpha\rangle = 0,  
\end{align}
for the flux-flipping operator, i.e., $j$ sites form the $z$ bond.

The remaining problem is the average value of $S^z_j$ that satisfies $[S^z_j, W_{p}]=0$ for all $p$. An $S^z_j$ operator satisfying such conditions depends on the lattice geometry and the type of topological phase, and thus must be considered on a case-by-case basis. Here, we restrict our discussion to the ($3,N_{\bm{a}_2}$) lattice geometry shown in Figs. \ref{Fig: two-plaquette} and the topological phase considered in Sec. \ref{sec: numerics}, where we assume that all redundant degeneracy is lifted by the perturbation terms, resulting in the twofold-degenerate ground state. As shown in Fig.~\ref{Fig: two-plaquette}, only $S_2^z=i\gamma_2\bar{\gamma}_2$ and $S_{N-1}^z= i\bar{\gamma}_{N-1}\gamma_{N-1}$ operators at edge sites do not form the $z$ bond, and thus, commute with all the $\mathbb{Z}_2$ fluxes. Since sites $2$ and $N-1$ remain after the single atoms are removed [see Fig.~\ref{Fig: main-concept} (b)], no Majorana zero modes exist there. Therefore, when the mode expansion is performed for $\gamma_2 $ and $\gamma_{N-1}$, these operators are expected not to include zero-energy components. Under this assumption, the operators $S^z_2$ and $S^z_{N-1}$ can be expressed as products of fermion operators with nonzero energy and localized Majorana operators. Thus, when condition (iii) is fulfilled, the ground state and excited states are orthogonal, leading to 
\begin{align}
&\langle \text{GS}_\alpha| S^z_{2} |\text{GS}_\alpha\rangle =\langle \text{GS}_\alpha| S^z_{N-1} |\text{GS}_\alpha\rangle = 0.
\end{align}

\section{Finite size effect and flux excitation energy}
\label{app:detail}

\begin{figure}[t]
    \centering
    \includegraphics[scale=0.25]{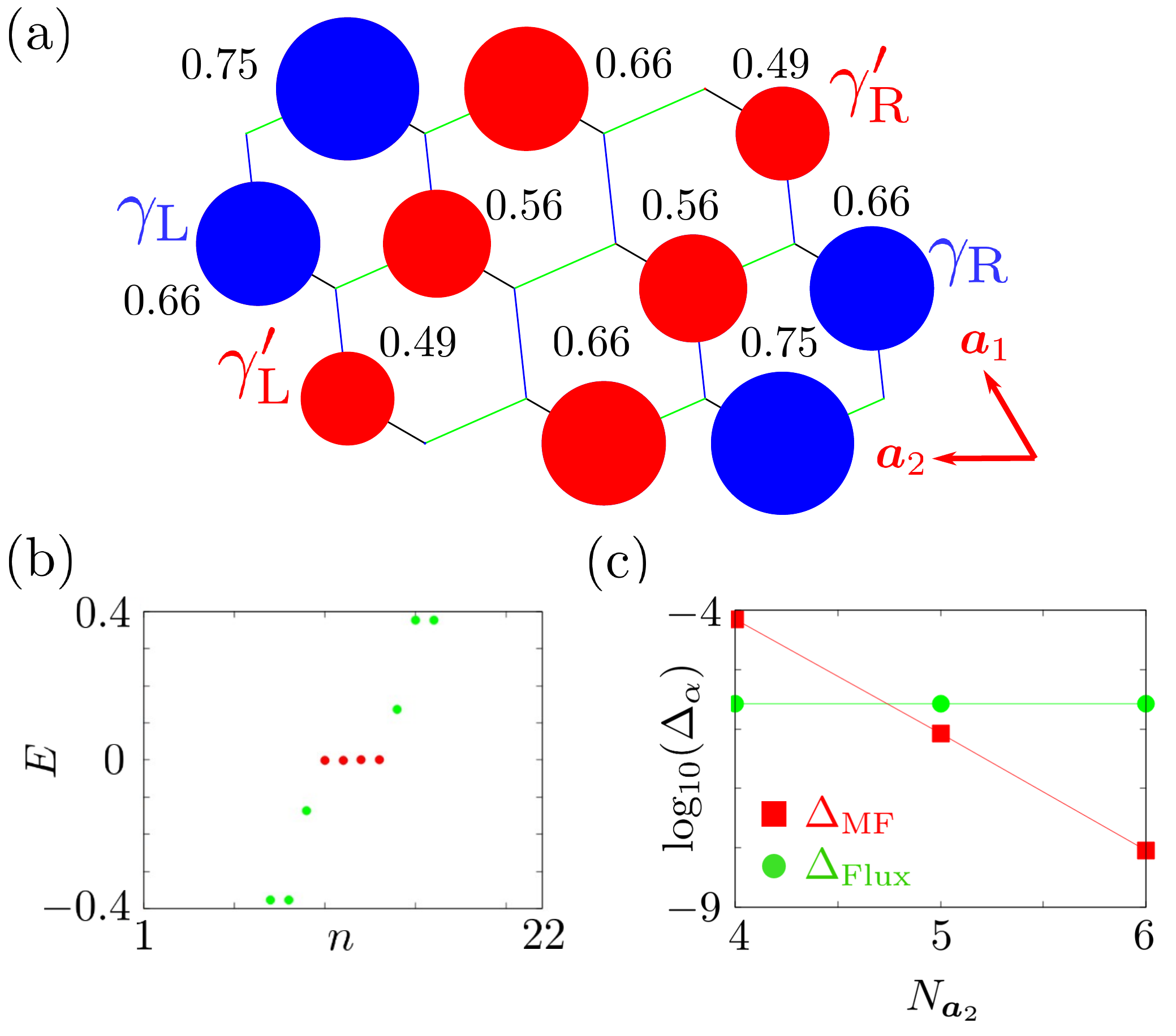}
    \caption{ Numerical results obtained by diagonalizing Eq.~(\ref{Majorana-Kitaev-2}) on the $(3,4)$ lattice geometry.  (a) Spatial distribution of the wave functions of Majorana zero modes $|u^{\alpha}_r|, \ (\alpha=\text{L,L',R,R'})$. Red and blue filled circles represent the probability density of $\gamma'_{\text{L/R}}$ and $\gamma_{\text{L/R}}$, respectively, with the circle size indicating the amplitude of the edge-state wave function at the corresponding site. (b) Energy spectrum of the itinerant Majorana fermions, where $n$ denotes the index of eigenvalues. Green dots correspond to bulk energy levels, while red dots indicate edge-state energy levels. (c) System-size dependence of $\Delta_{\text{MF}}$ and $\Delta_{\text{Flux}}$. The vertical axis shows $\text{log}_{10}(\Delta_\alpha)$ with $\alpha=$ MF, Flux.}
    \label{Fig: wavefunction}
\end{figure}

In this appendix, we examine the energy spectrum of the Hamiltonian (\ref{Majorana-Kitaev}) in the $(N_{\bm{a}_1},N_{\bm{a}_2})$ lattice geometry. 
For a given configuration $\{\eta_{\bm{r}}\}$, Eq.~(\ref{Majorana-Kitaev}) can be represented as 
\begin{align}
\nonumber
&H = \frac{1}{2} \bm{\gamma}^{\text{T}}H_{\eta}\bm{\gamma}, \\
&\bm{\gamma}=(\gamma_{1,1,A},\gamma_{2,1,A},...,\gamma_{N_{\bm{a}_1}-1,N_{\bm{a}_2},A},\gamma_{N_{\bm{a}_1},N_{\bm{a}_2},B}),
\label{Majorana-Kitaev-2}
\end{align}
with the matrix elements
\begin{align}
&[H_\eta]_{(i-1,j,A),(i,j,B)} =\frac{-iJ_x}{4}, \\
&[H_\eta]_{(i,j-1,A),(i,j,B)} =\frac{-iJ_y}{4}, \\
&[H_\eta]_{(i,j,A),(i,j,B)}  =\frac{-i\eta_{\bm{r}} J_z}{4}.
\end{align}
The other elements are zero.
Here, the edge sites $\gamma_{1,1,B}$ and $\gamma_{N_{\bm{a}_1},N_{\bm{a}_2},A}$ are removed so as to consider the lattice geometry shown in Fig.~\ref{Fig: main-concept} (b). Using a unitary matrix $U_\eta$, the Hamiltonian matrix is diagonalized as
\begin{align}
\nonumber
H &= \frac{1}{2} \bm{\gamma}^{\text{T}}U_\eta U^{\dagger}_\eta H_{\eta}U_\eta U^{\dagger}_\eta \bm{\gamma} \\ \nonumber
&= \sum_{m>0} \epsilon_m f^{\dagger}_{m,\eta}f_{m,\eta} + \sum_{m<0} \epsilon_m f^\dagger_{m,\eta}f_{m,\eta} \\ \nonumber
&= \sum_{m>0} [\epsilon_m f^{\dagger}_{m,\eta}f_{m,\eta}- \epsilon_m (1-f^{\dagger}_{m,\eta}f_{m,\eta})]  \\
&= 2\sum_{m>0} \epsilon_m f^{\dagger}_{m,\eta}f_{m,\eta} + E_{\text{GS}}, \label{eq: real-space-Hamiltonian} 
\end{align}
where 
$\bm{f}_{\eta} \equiv \frac{1}{\sqrt{2}}U_\eta \bm{\gamma}=(f_{1,\eta},f_{2,\eta},\cdots,f_{-1,\eta},f_{-2,\eta},\cdots)$ with $\{f^{\dagger}_{m,\eta},f_{m',\eta}\}=\delta_{mm'}$. $\epsilon_m$ are eigenvalues of $H_\eta$ satisfying $\epsilon_{-m} = -\epsilon_m$ due to the particle-hole symmetry. In the third line, we have used $f^{\dagger}_{m,\eta} =f_{-m,\eta}$. The ground-state energy is defined as $E_{\text{GS}} \equiv -\sum_{m>0} \epsilon_m$. Equation~(\ref{eq: real-space-Hamiltonian}) implicitly includes the Majorana zero modes with $\epsilon_0=0$. 

In the following, we numerically evaluate the quasiparticle energies $\epsilon_m$ and the unitary matrix $U_{\eta}$ in the $(3,N_{\bm{a}_2})$ lattice geometry. The coupling constants are fixed at $J_x=1.6$, $J_y=1.39$, and $J_z =0.01$, identical to those listed in Table~\ref{Tab: parameters}, while the configuration of $\eta_{\bm{r}}$ and the system size $N_{\bm{a}_2}>3$ are varied to examine the minimal flux excitation energy and its dependence on the lattice geometry.  Two configurations of $\{\eta_{\bm{r}}\}$ are considered: (i) $\eta_{\bm{r}} =+1$ for all $\bm{r}$, corresponding to the flux-free state, and (ii) one edge $\eta_{\bm{r}}=-1$ with all others $+1$, corresponding to the excited state with a single flux excitation.

We first analyze the flux-free states in the $(3,4)$ lattice geometry [see Fig.~\ref{Fig: wavefunction} (a)]. The corresponding energy spectrum $\epsilon_m$ is shown in Fig.~\ref{Fig: wavefunction} (b). As seen in the figure, four near-zero-energy states are present, corresponding to two Majorana zero modes localized at the left edge, $\gamma_{\rm L}$ and  $\gamma_{\rm L}'$, and two at the right edge, $\gamma_{\rm R}$ and  $\gamma_{\rm R}'$. The spatial distributions of these wave functions, extracted from the unitary matrix $U_{\eta}$, are displayed in Fig.~\ref{Fig: wavefunction} (a). Comparing with Eq.~(\ref{results3}), we confirm that $|\langle \text{GS}_\alpha|\sigma_N^y \sigma_1^y| \text{GS}_{\alpha}\rangle| \approx |u^{\text{L}}_N u^{\text{R}}_1|$. From Fig.~\ref{Fig: wavefunction} (b), the ground state energy and the bulk energy gap are estimated as $E_{\text{GS}} \approx -4.37684379994$ and $\Delta_{\text{bulk}} \approx 0.271$, respectively. 

Because of finite-size effects, the Majorana zero modes at the opposite edges hybridize, leading to small energy splittings away from zero. The energy difference between $\gamma'_{\text{L}}$ and $\gamma'_{\text{R}}$ is estimated as $\Delta_{\rm MF} \approx 7.0\times 10^{-5}$, while that between $\gamma_{\text{L}}$ and $\gamma_{\text{R}}$ is $\Delta_{\rm MF}' \approx 1.33 \times 10^{-9}$.

We next consider the excited state in the $(3,4)$ lattice geometry. As shown in Fig.~\ref{Fig: wavefunction} (c), the minimal one-flux excitation energy is estimated to be $\Delta_{\text{Flux}} \approx 2.66 \times 10^{-6}$. Since $\Delta_{\text{Flux}} < \Delta_{\text{MF}}$, this result appears to contradict the assumption that the flux-free state is the ground state. The discrepancy is attributed to the small system size and the resulting strong finite-size effects.

We expect that as the system size increases, i.e., as $N_{\bm{a}_2}$ becomes larger, $\Delta_{\rm MF}$ will become smaller than $\Delta_{\rm Flux}$. To confirm this expectation, we calculate $\Delta_{\rm MF}$ and $\Delta_{\rm Flux}$ for different values of $N_{\bm{a}_2}$. The results are presented in Fig.~\ref{Fig: wavefunction} (c), which clearly shows that $\Delta_{\rm MF} < \Delta_{\rm Flux}$ for $N_{\bm{a}_2} \ge 5$.

\bibliography{main}

\end{document}